\documentclass{aa}  
\usepackage{graphicx}
\usepackage{txfonts}
\usepackage{xcolor}
\usepackage{placeins}

\usepackage{hyperref}
\newcommand{\gaia}{\textit{Gaia}}

\begin{document}

   \title{First spectroscopic investigation of Anomalous Cepheid variables\thanks{Based on data obtained in the context of the ESO program 106.2129.001}}


   \author{V. Ripepi \inst{1}
          \and
          G. Catanzaro \inst{2}
          \and
          E. Trentin \inst{3,4,1}
         \and 
         O. Straniero \inst{5,6}
        \and
         A. Mucciarelli \inst{7,8} 
        \and
         M. Marconi \inst{1}
         \and
         A. Bhardwaj\inst{1} 
          \and    
          G. Fiorentino\inst{9}
          \and
          M. Monelli \inst{10,11,9}
          \and
          J. Storm\inst{3}
          \and
           G. De Somma \inst{1,12}
         \and       
           S. Leccia \inst{1}
          \and
          R. Molinaro \inst{1}
          \and 
          I. Musella \inst{1}
          \and
          T. Sicignano\inst{1}
          }

\institute{ INAF-Osservatorio Astronomico di Capodimonte, Salita Moiariello 16, 80131, Naples, Italy  \email{vincenzo.ripepi@inaf.it}
\and
INAF-Osservatorio Astrofisico di Catania, Via S.Sofia 78, 95123, Catania, Italy 
\and
Leibniz-Institut f\"ur Astrophysik Potsdam (AIP), An der Sternwarte 16, D-14482 Potsdam, Germany
\and
Institut für Physik und Astronomie, Universität Potsdam, Haus 28, Karl-Liebknecht-Str. 24/25, D-14476 Golm (Potsdam), Germany
             \and
Istituto Nazionale di Astrofisica – Osservatorio Astronomico d’Abruzzo, Via M. Maggini
s.n.c., 64100 Teramo, Italy
            \and
Istituto Nazionale di di Fisica Nucleare – sezione di Roma, Piazzale Aldo Moro 2, 00185
Roma, Italy
\and
Dipartimento di Fisica e Astronomia “Augusto Righi”, Alma Mater Studiorum, Università di Bologna, Via Gobetti 93/2, I-40129 Bologna, Italy
\and
INAF - Osservatorio di Astrofisica e Scienza dello Spazio di Bologna, Via Gobetti 93/3, I-40129 Bologna, Italy
\and
INAF - Osservatorio Astronomico di Roma, via Frascati 33, 00078
Monte Porzio Catone, Italy
\and
IAC- Instituto de Astrof\'isica de Canarias, Calle V\'ia Lactea s/n, E-38205 La Laguna, Tenerife, Spain
\and
Departmento de Astrof\'isica, Universidad de La Laguna, E-38206 La Laguna, Tenerife, Spain
\and
Istituto Nazionale di Fisica Nucleare (INFN)-Sez. di Napoli, Via Cinthia, 80126 Napoli, Italy
}

   \date{Received September 15, 1996; accepted March 16, 1997}

 
  \abstract
   {Anomalous Cepheids (ACEPs) are intermediate mass metal-poor pulsators mostly discovered in dwarf galaxies of the Local Group. However, recent Galactic surveys, including the Gaia Data Release 3, found a few hundreds of ACEPs in the Milky Way. Their origin is not well understood.}
   {We aim to investigate the origin and evolution of Galactic ACEPs by studying for the first time the chemical composition of their atmospheres.}
   {We used UVES@VLT to obtain high-resolution spectra for a sample of 9 ACEPs belonging to the Galactic halo. We derived the abundances of 12 elements, including C, Na, Mg, Si, Ca, Sc, Ti, Cr, Fe, Ni, Y, and Ba. We complemented these data with literature abundances from high-resolution spectroscopy for an additional three ACEPs which were previously incorrectly classified as type II Cepheids, thus increasing the sample to a  total of 12 stars. }
   {All the investigated ACEPs have an iron abundance [Fe/H]$<-1.5$ dex as expected from theoretical predictions for these pulsators. The abundance ratios of the different elements to iron show that the ACEP's chemical composition is generally consistent with that of the Galactic halo field stars, with the exception of the Sodium, which is found overabundant in 9 out of the 11 ACEPs where it was measured, in close similarity with second-generation stars in the Galactic Globular Clusters. The same comparison with dwarf and ultra-faint satellites of the Milky Way reveals more differences than similarities so it is unlikely that the bulk of Galactic ACEPs originated in such a kind of galaxies which subsequently dissolved in the Galactic halo. 
   The principal finding of this work is the unexpected overabundance of Sodium in ACEPs. We explored several hypotheses to explain this feature, finding that the most promising scenario is the evolution of low-mass stars in a binary system with either mass transfer or merging. Detailed modelling is needed to confirm this hypothesis. }
   {}


   \maketitle
%

\section{Introduction}

Anomalous Cepheids (ACEPs) are short-period pulsating variables with periods approximately between 0.4 and 2.5 days. Historically they have been called Anomalous because at a fixed period, they are brighter than the BL Herculis (BLHER) subclass of type II Cepheids. 
Similar to their more famous kins, RR Lyrae and classical Cepheid variables, they pulsate in the fundamental (ACEP\_F) and first-overtone (ACEP\_1O) mode, while only one mixed-mode ACEP has been found so far in the Large Magellanic Cloud \citep[LMC][]{Sos2020}. At fixed period, the ACEPs have light curves similar to that of RR Lyrae, short-period classical Cepheids as well as BLHER, so it is often difficult to distinguish ACEPs from the pulsators mentioned above when the stars are not at the same distances, such as Local Group galaxies  
\citep[see e.g.][]{Plachy2021,Ripepi2023}. 
Similar to classical Cepheids, ACEPs follow distinct period-luminosity (PL) relations for the different pulsation modes \citep[e.g.][and references therein]{Ripepi2014-ACEP,Sos2015-ACEP-MC,Ngeow2022}. The tightness of these relations, especially in the near-infrared (NIR) bands, make ACEPs good standard candles for Local Group galaxies \citep[e.g.][]{Ripepi2014-ACEP}. 

In the colour-magnitude diagram (CMD), the ACEPs are located in the classical instability strip, roughly at the same colours as that of RR Lyrae variables but at magnitudes about 0.75 to 2.2 mag brighter.  
From the stellar evolutionary point of view, the ACEPs are metal-poor ([Fe/H]$<-1.5$ dex), intermediate-mass stars (1.3--2.5 M$_{\odot}$), in their Helium core burning phase which has ignited Helium combustion under partial electron-degeneracy conditions \citep[][]{Renzini1977, CastellaniScilla1995, Bono1997, Caputo2004, Fiorentino2006,FiorentinoMonelli2012,Monelli2022}.
Stars with larger masses ignite the Helium quiescently and become short-period classical Cepheids, while for too metal-rich intermediate-mass stars the excursion to the blue during core-Helium burning is not sufficient to enter the classical instability strip \citep[see e.g.][]{Caputo2004,Marconi2004,Monelli2022}.

ACEPs have been discovered long ago in dwarf spheroidal galaxies (dSph) of the Local Group Sculptor \citep{Thackeray1950} and Draco \citep{Baade1961}. Since then they have been discovered in many dwarf galaxies of the Local Group displaying very different properties. Indeed, ACEPs have been found in the Milky Way (MW) satellites showing typically purely old and metal-poor stellar populations (ages larger than 10 Gyr) such as the dSph Draco, Leo\,II, Sculptor, Sextans, Ursa Minor \citep{ZinnSearle1976,Siegel2000,Smith1986,Mateo1995,Nemec1988}; the ultra-faint dwarfs (UFDs) CVnI, Hercules, Eridanus\,II \citep{Kuehn2008,Musella2012,Clara2021}; the isolated dwarfs Cetus and Tucana \citep{Monelli2012-Cetus,Bernard-2009}; and the two galactic globular clusters (GGCs) NGC\,5466 and M92 \citep{ZinnDahn1976,Ngeow2022}. ACEP variables are also hosted by dSph galaxies showing large intermediate age (1-6 Gyr) populations such as Fornax and Carina \citep{Bersier2002,Coppola2015}; the peculiar gas-rich UFD Leo\,T \citep{Clementini2012} and the gas-rich dwarf Phoenix \citep{Gallart2004,Ordonez-2014}.  
In other cases, the ACEPs can be confused with metal-poor short-period classical Cepheids \citep[e.g. NGC\,6822 and Leo\,A][]{Hoessel1994,Baldacci2005,Dolphin2002,Bernard2013}. Finally, both ACEPs and classical Cepheids have been found in Leo\,I \citep{Stetson2014} and in the most massive satellites of the MW, namely the Large and Small Magellanic Clouds \citep[LMC and SMC][]{Sos2015-ACEP-MC}. In more detail, the LMC and SMC host a large intermediate-age population which, based on current LMC/SMC age-metallicity relations \citep[see e.g.][for the LMC and SMC, respectively]{Gatto2022,Piatti2015} at the expected age of ACEPs ($\sim$1--6 Gyr) would show a metallicity much larger than that predicted to enter the instability strip for ACEP pulsation. 

As mentioned above, there is a consensus on the mass range spanned by ACEPs. However, the presence of these pulsators in a variety of stellar systems hosting pure old to intermediate-age populations poses serious uncertainties on the origin of the ACEPs progenitors.   
To date, two channels for ACEP formation have been proposed in the literature. The first scenario is the evolution of single stars as a result of a star formation event that occurred about 1 to 6 Gyr ago in a metal-poor environment \citep[e.g.][]{NorrisZinn1975,ZinnSearle1976,CastellaniScilla1995,Caputo1998,Caputo2004}. This channel can explain the presence of ACEPs in (metal-poor) dSph with extended star formation history which produced an intermediate-age population (e.g. Fornax and Carina). 
The second channel predicts that ACEPs are the evolved descendant of Blue Straggler (BS) stars\footnote{i.e., a main-sequence star brighter, bluer and more massive than normal halo stars close to the turn-off point.}, formed via mass-transfer in binary systems \citep[][]{McCrea1964,Renzini1977,Wheeler1979,Sills-2009,Gautschy2017}. This mechanism would explain not only the presence of ACEPs in the old and metal-poor population but also in more metal-rich host galaxies, such as the LMC \citep[][]{Gautschy2017}. 

This scheme seems to be supported by the study of the specific frequency of ACEPs (frequency of ACEPs per unit of luminosity) carried out by \citet{Monelli2022} in a sample of 26 Local Group galaxies divided into fast and slow systems. The former generated the great majority (or all) of the stars at remote times, namely $>$10 Gyr ago. At the same time, the latter produced part of their stars at old epochs but exhibited a significant fraction of intermediate-age (or even young) populations. \citet{Monelli2022} found that, at fixed luminosity, slow systems tend to produce more ACEPs than fast ones. This can be explained by hypothesising that slow galaxies can produce ACEPs through both the single and binary star mechanisms.
However, not all the observational evidence fits into this scenario. For example, the lack of ACEPs in (metal-poor) GGCs having large populations of BSs (except the mentioned NGC\,5466 and M92) cannot be easily reconciled with the above scenario.

   \begin{figure}
   \centering
   \includegraphics[width=\hsize]{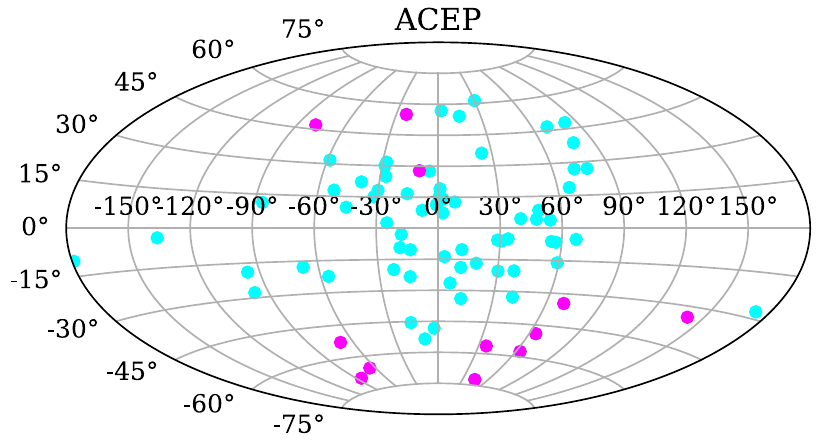}
      \caption{Position on the sky in Galactic coordinates for the nine objects studied in this work (magenta-filled circles in the south hemisphere). The three complementary stars taken from the literature are also shown (magenta-filled circles in the north hemisphere). For comparison, light-blue dots display the position of all the ACEPs in the \gaia\ DR3 catalogue.
       }
         \label{fig:aitoff}
   \end{figure}

   \begin{figure}
   \centering
   \includegraphics[width=\hsize]{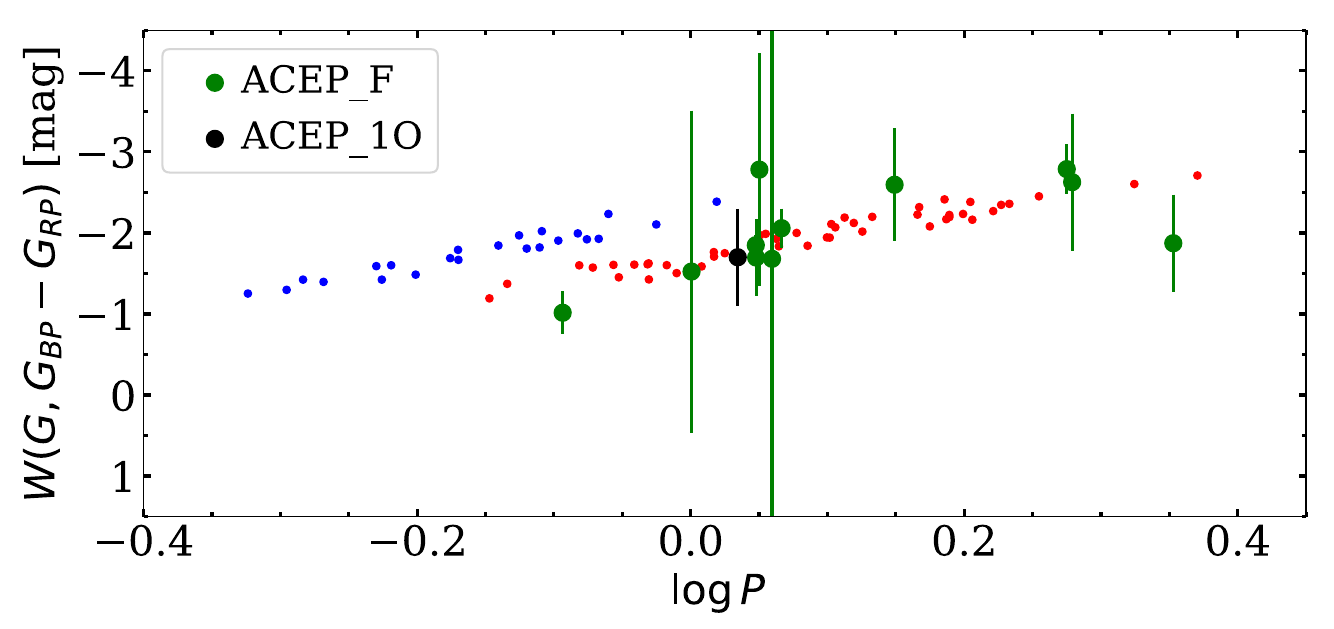}
      \caption{Period-Wesenheit relation in the \gaia\ bands for the 11 ACEP\_Fs (filled green circles) and 1 ACEP\_1O (black filled circle, $\log P$ shifted by -0.01 to avoid overlap with F pulsators) analysed in this work in comparison with analogue data for the LMC (red and blue dots represent ACEP\_Fs and ACEP\_1Os, respectively). To obtain the absolute Wesenheit magnitude for the LMC stars we adopted the distance modulus by \citet{Pietrzynski2019}.
      }
         \label{fig:pw}
   \end{figure}

\begin{table*}
\caption{Basic data for the program stars and for the objects taken from the literature.} 
\label{tab:basicData}
\footnotesize\setlength{\tabcolsep}{2pt}
\begin{tabular}{lrlrrcllcrc}
\hline
\hline
  \multicolumn{1}{c}{Name} &
  \multicolumn{1}{c}{source\_id} &
  \multicolumn{1}{c}{DR} &
  \multicolumn{1}{c}{RA} &
  \multicolumn{1}{c}{Dec} &
  \multicolumn{1}{c}{Mode} &
  \multicolumn{1}{c}{Period} &
  \multicolumn{1}{c}{$V$} &
  \multicolumn{1}{c}{$\varpi/\sigma \varpi$}  &
   \multicolumn{1}{c}{$D$} & 
    \multicolumn{1}{c}{$E(B-V)$}\\

  \multicolumn{1}{c}{} &
  \multicolumn{1}{c}{} &
  \multicolumn{1}{c}{} &
  \multicolumn{1}{c}{deg} &
  \multicolumn{1}{c}{deg} &
  \multicolumn{1}{c}{} &
  \multicolumn{1}{c}{days} &
  \multicolumn{1}{c}{mag} &
  \multicolumn{1}{c}{} &
    \multicolumn{1}{c}{kpc} &
      \multicolumn{1}{c}{mag} \\

\hline
  DR2 6498717390695909376     & 6498717390695909376 & DR2 & 351.00495 & $-$57.04665  &  F & 0.806120 & 13.914 & 8.6        &   7.1$\pm$0.6  &  0.014 \\
  DR2 2976160827140900096     & 2976160827140900096 & DR3 & 73.77906  & $-$19.18685  &  F & 1.001443 & 15.593 & 1.2	     &  15.3$\pm$1.3  &  0.043 \\
  HE 2324-1255                & 2409340558427973248 & DR3 & 351.82896 & $-$12.64747  & 1O & 1.107658 & 13.849 & 3.8	     &  10.1$\pm$0.8  &  0.034 \\
  V716 Oph$^{a}$              & 4352629469629411840 & DR3 & 247.70599 & $-$5.50547   &  F & 1.115890 & 12.033 & 17.8       &   2.3$\pm$0.2  &  0.438 \\
  HE 0114-5929                & 4909045712639882496 & DR3 & 19.20812  & $-$59.23468  &  F & 1.116570 & 14.321 & 4.9	     &   9.7$\pm$0.8  &  0.025 \\
  DF Hyi                      & 4698416118397803776 & DR3 & 25.20499  & $-$67.49499  &  F & 1.122586 & 14.188 & 1.6	     &  10.3$\pm$0.9  &  0.026 \\
  SHM2017 J000.07389-10.22146 & 2422853521974230400 & DR3 & 0.07391   & $-$10.22146  &  F & 1.146828 & 15.073 & 0.6	     &  16.4$\pm$1.4  &  0.036 \\
  BF Ser$^{a}$                & 1208200864738741376 & DR3 & 229.11868 &   16.44430   &  F & 1.165421 & 11.986 & 9.6        &   3.9$\pm$0.3  &  0.033 \\
  EK Del$^{b}$                & 1400474455952839168 & DR3 & 240.93063 &   50.22592   &  F & 1.409596 & 14.081 & 3.3	     &  10.4$\pm$0.9  &  0.024 \\
  OGLE GAL-ACEP-006           & 4627678075752483584 & DR3 & 65.96439  & $-$76.91187  &  F & 1.883580 & 13.015 & 7.5	     &   6.2$\pm$0.5  &  0.094 \\
  CRTS J003041.3-441620       & 4980356188527065472 & DR3 & 7.67244   & $-$44.27295  &  F & 1.900711 & 13.363 & 2.7	     &   9.4$\pm$0.8  &  0.010 \\
  DR2 2648605764784426624     & 2648605764784426624 & DR2 & 343.43880 & $-$2.81562   &  F & 2.253570 & 14.077 & 3.8        &  11.7$\pm$1.0  &  0.074 \\
\hline
\end{tabular}
\tablefoot{Name is the star identifications on Simbad; source\_id is the $Gaia$ identifier; DR is the $Gaia$ data release; RA and Dec are the equatorial coordinates (J2000); Mode is the pulsation mode; Period is the period of pulsation; $V$ is the visual magnitude; $\varpi/\sigma \varpi$ is the relative error on the $Gaia$ parallaxes; D is the distance calculated as described in Appendix~\ref{AppendixC};  E(B-V) is the reddening calculated from the \citet{Schlafly2011} maps using the {\tt dustmaps} package \citep{Green2018}; a=\citet{Kovtyukh2018};  
b=\citet{Luck2011}} 
\end{table*}

Concerning the MW, hundreds of ACEPs have been discovered in recent years in almost all the Galaxy components, namely, bulge, disk and halo, thanks to wide surveys such as OGLE, Catalina sky survey, and \gaia\ mission \citep[][]{Udalski2018,Sos2020,Drake2014,Torrealba2015,Clementini2016,Clementini2019,Ripepi2019,Ripepi2023}. Given the uncertain genesis of ACEPs, these findings naturally raise the question about the origin of these objects. Were they born in situ, or were they accreted during past merging episodes with dSph and UFD galaxies? Can we discern whether they result from binary or single-star evolution?

The scope of this paper is to investigate these subjects by using the chemical tagging of a sample of Galactic ACEPs through high-resolution spectroscopy. We stress that no detailed chemical analyses of ACEPs are available so far.

\section{Sample selection, observations, data reduction, and abundance derivation}

In this section, we describe the sample selection, the observation and data reduction. We also explain how the chemical abundances have been obtained.

   \begin{figure*}
   \centering
   \includegraphics[width=\hsize]{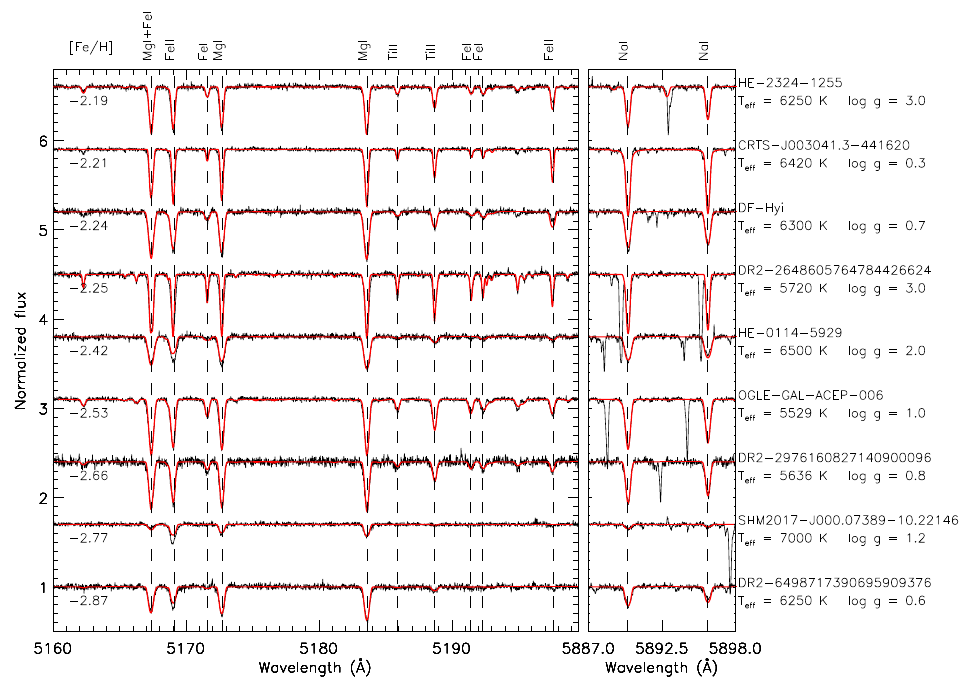}
      \caption{Excerpt from our UVES spectra for the nine ACEPs studied in this work. 
       Black and red lines show the data and the best-fitting synthetic spectra, respectively. 
       Some lines are labelled in the figure as well as the iron abundance. In the right panel, we show the fit of the \ion{Na}{i} doublet at 5889-5895 {\AA}. It is possible to note the presence of interstellar sodium lines in 4 out of 9 stars, for which we accurately fit only the line profile of the stellar sodium.
       }
         \label{fig:spectra}
   \end{figure*}

\subsection{Sample selection}

The program stars are listed in Table~\ref{tab:basicData}. They were selected based on the sample of Cepheid data in \gaia\ Data Release 2 \citep[DR2][]{Gaia2016,Gaia2018,Clementini2019} as re-classified by \citet{Ripepi2019}. Their classification was further refined by means of the \gaia\ Early Data Release 3 \citep[DR3][]{Gaia2021} parallaxes \citep[in particular for the stars HE\,2324-1255 and DR2\,2648605764784426624, which were not classified as ACEPs in][]{Ripepi2019}.       
To increase the sample, we cross-matched the \gaia\ DR3 catalogue of ACEPs with literature metallicity observations based on high-resolution spectroscopy. We found three stars, namely EK\,Del, BF\,Ser and V716\,Oph, among which EK Del was erroneously classified as classical Cepheid while BF\,Ser and V716\,Oph were usually considered as type II Cepheids \citep[e.g.][]{Kovtyukh2018}, but their ACEP nature was suspected by \citep{Drake2014} and later confirmed by \citep{Jurkovic2018}. For these stars, we adopted the chemical abundances of \citet{Kovtyukh2018} for BF\,Ser and V716\,Oph, while for EK\,Del we used the results by \citet{Luck2011}. The general characteristics of these stars are listed in Table~\ref{tab:basicData} along with those of the program stars. The position in the Galaxy of all the 12 objects discussed in this work is shown in Fig.~\ref{fig:aitoff}. We can safely assume that they are all located in the Galactic Halo. 

To further assess the classification as ACEPs of the 12 stars considered here, we report in Fig.~\ref{fig:lc} their light curves in the \gaia\ bands. All the data were taken from DR3, with the exceptions of stars DR2\,6498717390695909376 and DR2\,2648605764784426624, which are present in DR2 only \citep[see][for explanations]{Ripepi2023}. The shapes of the light curves vary as the period increases from about one to about 2.25 days. In particular, EK\,Del, BF\,Ser and V716\,Oph show asymmetric 
light curves with bumps and humps that are typical of ACEPs of the corresponding periods.

Figure~\ref{fig:pw} shows the period-Wesenheit\footnote{The Wesenheit magnitudes are reddening-free by construction \citep{Madore1982}} (PW) relation in the \gaia\ bands \citep[defined as $G-1.90 \times (G_{BP}-G_{RP})$, see ][]{Ripepi2019} for the 12 ACEPs analyzed in this work in comparison with the ACEPs from the LMC. For the Galactic ACEPs we obtained the photometry from \gaia\ Data Release 3 \citep[][]{GaiaVallenari,Ripepi2023}, while the absolute Wesenheit magnitudes were calculated by inverting the \gaia\ parallaxes after correcting for the zero-point offset according to \citet{Lindegren2021}\footnote{We are aware of the incorrectness of this procedure for parallaxes with relative error larger than about 10-15\%, indeed the figure is only for illustrative purposes.}. Similarly, for the LMC, the photometry is from \gaia\ DR3, while absolute Wesenheit magnitudes have been derived by adopting a distance modulus of 18.477$\pm$0.026 mag \citep{Pietrzynski2019}.
The PW of the MW ACEPs investigated in this article are compatible with the expected position as traced by the well-known LMC objects, thus confirming the correctness of the classification\footnote{The only ACEP\_1O star, namely HE 2324-1255, is an exception in the sense that its position in Fig.~\ref{fig:pw} is more in agreement with F-mode pulsators than with 1O-mode ones. However, the analysis of the light curves shown in Fig.~\ref{fig:lc} leaves little doubt about its classification as an overtone pulsator. }.  

\begin{table*}
\caption{Atmospheric parameters for the program stars.}
\label{tab:atm_err}
\begin{tabular}{lccccrcc}
\hline
\hline
  \multicolumn{1}{c}{Name} &
  \multicolumn{1}{c}{T$_{eff}$} &
  \multicolumn{1}{c}{log g} &
  \multicolumn{1}{c}{v$_{br}$} &
  \multicolumn{1}{c}{$\xi$} &
  \multicolumn{1}{c}{v$_{rad}$}  &
\multicolumn{1}{c}{Phase}  &
\multicolumn{1}{c}{T$_{eff}^{Phot}$} \\
  \multicolumn{1}{c}{} &
  \multicolumn{1}{c}{K} &
  \multicolumn{1}{c}{} &
  \multicolumn{1}{c}{km s$^{-1}$} &
  \multicolumn{1}{c}{km s$^{-1}$} &
  \multicolumn{1}{c}{km s$^{-1}$} &
    \multicolumn{1}{c}{} &
  \multicolumn{1}{c}{K} \\
\hline
 DR2 6498717390695909376       &  6250 $\pm$ 220 & 2.0 $\pm$ 0.2 & 17 $\pm$ 2 & 1.0 $\pm$ 0.3  &  137.6 $\pm$ 1.1 &  0.68  &  6250 $\pm$ 105 \\
 DR2 2976160827140900096       &  5750 $\pm$ 200 & 1.0 $\pm$ 0.2 & 13 $\pm$ 1 & 3.2 $\pm$ 0.6  &  209.5 $\pm$ 0.8 &  0.56  &  5830 $\pm$ 100 \\
 HE 2324-1255                  &  6250 $\pm$ 180 & 0.7 $\pm$ 0.1 & 12 $\pm$ 1 & 0.6 $\pm$ 0.2  & -137.2 $\pm$ 0.4 &  0.22  &  6630 $\pm$ 130 \\
 HE 0114-5929                  &  6500 $\pm$ 200 & 3.0 $\pm$ 0.5 & 20 $\pm$ 2 & 1.0 $\pm$ 0.3  &  124.1 $\pm$ 0.8 &  0.85  &  6300 $\pm$ 210 \\
 DF Hyi                        &  6300 $\pm$ 270 & 1.2 $\pm$ 0.1 & 16 $\pm$ 2 & 2.3 $\pm$ 0.5  &  243.7 $\pm$ 0.6 &  0.86  &  6120 $\pm$ 115 \\
 SHM2017 J000.07389-10.22146   &  7000 $\pm$ 250 & 3.0 $\pm$ 0.5 & 18 $\pm$ 2 & 1.0 $\pm$ 0.3  & -367.9 $\pm$ 1.0 &  0.98  &  8370 $\pm$ 130 \\
 OGLE GAL-ACEP-006             &  5500 $\pm$ 150 & 0.3 $\pm$ 0.1 & 13 $\pm$ 1 & 3.0 $\pm$ 0.7  &   97.1 $\pm$ 0.5 &  0.70  &  5770 $\pm$ 130 \\
 CRTS J003041.3-441620         &  6420 $\pm$ 200 & 0.6 $\pm$ 0.1 &  8 $\pm$ 1 & 0.0 $\pm$ 0.3  &   47.7 $\pm$ 0.4 &  0.18  &  6520 $\pm$ 100 \\
 DR2 2648605764784426624       &  5750 $\pm$ 180 & 0.8 $\pm$ 0.2 &  6 $\pm$ 1 & 2.2 $\pm$ 0.6  &   47.5 $\pm$ 0.4 &  0.19  &  5950 $\pm$ 130 \\
\hline
\end{tabular}
\end{table*}

\subsection{Observations and data reduction}

The spectroscopic observations analyzed here have been acquired with the ESO (European Southern Observatory) UVES (Ultraviolet and Visual Echelle Spectrograph\footnote{https://www.eso.org/sci/facilities/paranal/instruments/uves.html}) instrument, which is operated at the Unit Telescope 2 of VLT (Very Large Telescope), placed at Paranal (Chile). The data were obtained on December, 12-20 2020 in the context of the proposal P106.2129.001 which was mostly devoted to observing a large sample of Classical Cepheids which we published in \citet{Trentin2023}.
We adopted the red arm and the grism CD\#3 which has a central wavelength at 5800 {\AA} and covers the wavelength interval 4760--6840 {\AA}. We choose the 1 arcsec slit, which gives a dispersion of R$\sim$47,000. 
As the instrumental setup was identical, we refer the reader to \citet{Trentin2023} for full details on the data reduction steps that led to the production of the fully calibrated normalised spectra. 

Figure~\ref{fig:lc} shows the phases at which each star was observed \footnote{For the literature stars, the phases were taken from the source papers. BF\,Ser and V716\,Oph were observed repeatedly at different phases, only one phase is shown in the figure for homogeneity}. The majority of the targets have been observed in the quiescent descending branch. For a few stars, namely  HE\,0114-5929, DF\,Hyi and SHM2017\,J000.07389-10.22146, the observations fell instead on the fast-rising branch or at maximum light. Albeit it is preferable to avoid such phases of the pulsation cycle, we do not expect that our abundance analysis can be significantly affected \citep[see e.g. section 7 in ][for a discussion about this point in the context of RR Lyrae pulsators]{For2011}.  

\begin{table*}
\caption{Abundances for the stars analysed in this work. All the data is expressed as [X/H] (dex).}
\label{tab:abundances}
\footnotesize\setlength{\tabcolsep}{3pt}
\begin{tabular}{lccccccc}
\hline
\hline
  Star                       &         C            &       Na            &        Mg           &       Si            &        Ca            &          Sc           &          Ti      \\
\hline                                                                                                                                           
DR2 6498717390695909376      &            --        & $-$2.31 $\pm$ 0.11  & $-$2.33 $\pm$ 0.30  &          --         & $-$2.28 $\pm$  0.11  &           --          &           --         \\ 
DR2 2976160827140900096      &            --        & $-$2.18 $\pm$ 0.11  & $-$2.37 $\pm$ 0.15  &          --         & $-$2.02 $\pm$  0.17  &  $-$2.82 $\pm$  0.06  &  $-$2.46 $\pm$  0.17 \\
HE 2324-1255                 &            --        & $-$1.44 $\pm$ 0.11  & $-$1.89 $\pm$ 0.08  & $-$2.22 $\pm$ 0.09  & $-$1.78 $\pm$  0.17  &  $-$2.50 $\pm$  0.16  &  $-$2.30 $\pm$  0.15 \\
V716 Oph$^{a}$               &  $-$1.21 $\pm$ 0.20  & $-$1.28 $\pm$ 0.20  & $-$1.43 $\pm$ 0.20  & $-$1.50 $\pm$ 0.20  & $-$1.11 $\pm$  0.20  &  $-$1.41 $\pm$  0.20  &  $-$1.15 $\pm$  0.20 \\
HE 0114-5929                 &           --         & $-$1.31 $\pm$ 0.11  & $-$2.15 $\pm$ 0.18  &          --         & $-$2.01 $\pm$  0.18  &           --          &           --         \\
DF Hyi                       &           --         & $-$1.25 $\pm$ 0.11  & $-$1.78 $\pm$ 0.12  & $-$2.35 $\pm$ 0.15  & $-$1.75 $\pm$  0.13  &  $-$2.54 $\pm$  0.14  &  $-$2.28 $\pm$  0.10 \\
SHM2017 J000.07389-10.22146  &           --         & $-$3.04 $\pm$ 0.04  & $-$3.19 $\pm$ 0.06  & $-$1.32 $\pm$ 0.05  &          --          &           --          &           --         \\
BF Ser$^{a}$                 &  $-$1.95 $\pm$ 0.20  & $-$2.28 $\pm$ 0.20  & $-$1.77 $\pm$ 0.20  &          --         & $-$1.33 $\pm$  0.20  &  $-$1.88 $\pm$  0.20  &  $-$1.67 $\pm$  0.20 \\
EK Del$^{b}$                 &  $-$1.88 $\pm$ 0.20  &          --         & $-$1.53 $\pm$ 0.20  & $-$1.17 $\pm$ 0.20  & $-$1.24 $\pm$  0.20  &           --          &  $-$1.02 $\pm$  0.20 \\
OGLE GAL-ACEP-006            &           --         & $-$1.67 $\pm$ 0.20  & $-$2.15 $\pm$ 0.16  & $-$2.03 $\pm$ 0.10  & $-$1.81 $\pm$  0.12  &  $-$2.52 $\pm$  0.13  &  $-$2.33 $\pm$  0.16 \\
CRTS J003041.3-441620        &  $-$1.79 $\pm$ 0.06  & $-$0.94 $\pm$ 0.11  & $-$1.83 $\pm$ 0.13  & $-$2.25 $\pm$ 0.06  & $-$1.73 $\pm$  0.12  &  $-$2.42 $\pm$  0.15  &  $-$2.27 $\pm$  0.16 \\
DR2 2648605764784426624      &           --         & $-$1.53 $\pm$ 0.16  & $-$2.12 $\pm$ 0.17  & $-$1.67 $\pm$ 0.15  & $-$1.75 $\pm$  0.16  &  $-$2.19 $\pm$  0.12  &  $-$1.94 $\pm$  0.11 \\
\hline
 \multicolumn{8}{c}{Continued} \\
\hline  Star                         &          Cr        &             Fe       &          Ni       &         Y        &         Ba       \\
\hline                                                                                                            
DR2 6498717390695909376        &   $-$2.97 $\pm$  0.16  &  $-$2.87 $\pm$ 0.15  &            --         &             --        &           --          \\ 
DR2 2976160827140900096        &   $-$2.56 $\pm$  0.08  &  $-$2.66 $\pm$ 0.14  &            --         &             --        &  $-$2.89 $\pm$  0.13  \\
HE 2324-1255                   &   $-$2.28 $\pm$  0.12  &  $-$2.19 $\pm$ 0.17  &            --         &             --        &  $-$3.46 $\pm$  0.13  \\
V716 Oph$^{a}$                 &   $-$1.49 $\pm$  0.20  &  $-$1.64 $\pm$ 0.20  &  $-$1.32 $\pm$  0.20  &  $-$1.46 $\pm$  0.20  &  $-$1.69 $\pm$  0.20  \\
HE 0114-5929                   &   $-$2.41 $\pm$  0.16  &  $-$2.42 $\pm$ 0.10  &           --          &           --          &           --          \\
DF Hyi                         &   $-$2.02 $\pm$  0.11  &  $-$2.24 $\pm$ 0.12  &  $-$2.03 $\pm$  0.07  &  $-$2.37 $\pm$  0.16  &  $-$2.63 $\pm$  0.16  \\
SHM2017 J000.07389-10.22146    &            --          &  $-$2.77 $\pm$ 0.10  &           --          &           --          &            --         \\
BF Ser$^{a}$                   &   $-$1.82 $\pm$  0.20  &  $-$2.08 $\pm$ 0.20  &           --          &  $-$1.80 $\pm$  0.20  &  $-$1.88 $\pm$  0.20  \\
EK Del$^{b}$                   &   $-$1.64 $\pm$  0.20  &  $-$1.54 $\pm$ 0.20  &  $-$1.28 $\pm$  0.20  &  $-$1.40 $\pm$  0.20  &           --          \\
OGLE GAL-ACEP-006              &   $-$2.48 $\pm$  0.16  &  $-$2.53 $\pm$ 0.11  &           --          &  $-$3.07 $\pm$  0.11  &  $-$2.61 $\pm$  0.13  \\
CRTS J003041.3-441620          &   $-$2.20 $\pm$  0.16  &  $-$2.21 $\pm$ 0.09  &  $-$2.01 $\pm$  0.11  &  $-$2.42 $\pm$  0.16  &  $-$1.77 $\pm$  0.22  \\
DR2 2648605764784426624        &   $-$2.10 $\pm$  0.15  &  $-$2.25 $\pm$ 0.15  &  $-$2.25 $\pm$  0.17  &  $-$2.68 $\pm$  0.06  &  $-$2.35 $\pm$  0.13  \\
\hline 
\hline
\end{tabular}
\tablefoot{a=\citet{Kovtyukh2018}\\
b=\citet{Luck2011}} 
\end{table*}

\subsection{Atmospheric parameters}

The chemical analysis has been performed essentially in the same way as described in \citet{Trentin2023}. Here we briefly recall the main steps of the procedure. 
First, we estimated the atmospheric parameters, namely the effective temperature (T$_{\rm eff}$), surface gravity (log {\itshape g}), microturbulent velocity ($\xi$), and line broadening parameter (v$_{\rm br}$), i.e. the combined effects of macroturbulence and rotational velocity.

Given the low metallicity of these stars, the number of measurable spectral lines of \ion{Fe}{I} in their spectrum is extremely limited. For this reason, in order to determine the atmospheric parameters necessary for chemical analysis, we had to proceed in two ways, depending on the number of detected iron lines.

For the stars where a sufficient number of spectral lines could be measured, we applied the method of excitation equilibrium. This method ensures that there is no residual correlation between the iron abundance and the excitation potential of the neutral iron lines \citep[see e.g.,][]{mucciarelli2020}. We estimated the microturbulence by requiring that the slope of [Fe/H] as a function of equivalent widths (EWs) is zero. To achieve this, we initially measured the equivalent widths of a sample of \ion{Fe}{I} lines using a semi-automatic custom routine in {\it IDL}\footnote{IDL (Interactive Data Language) is a registered trademark of L3HARRIS Geospatial}. The conversion of EWs to abundance was performed using the WIDTH9 code \citep{kurucz1981solar} applied to the corresponding atmospheric model calculated using ATLAS9 \citep{kurucz1993new}. In this calculation, we did not consider the influence of $\log$ {\itshape g} since neutral iron lines are insensitive to it. Next, we estimated the surface gravities iteratively by imposing the ionisation equilibrium between \ion{Fe}{I} and \ion{Fe}{II}.

The sample of \ion{Fe}{I} and \ion{Fe}{II} lines was extracted from the line list published by \citet{romaniello2008}.

For the remaining objects (HE 0114-5929 and SHM2017 J000.07389-10.22146), we estimated the temperature by fitting the H$\beta$ profile. In practice, we minimized the difference between the observed and synthetic spectra using the $\chi^2$ as a goodness-of-fit parameter. The synthetic profile was generated in three steps, as described in the next section. It is known from the literature that the core of Balmer lines is not reproduced by 1D-LTE models, for which 3D non-LTE models would be necessary. 
The global effects of non-LTE on Balmer lines have been extensively studied by \citet{mashonkina2008} and \citet{amarsi2018}. Both studies conclude that neglecting non-LTE effects in metal-poor stars leads to an underestimate of the effective temperature by approximately 150 K, which we have taken into account in the total error estimation. We determined the projected rotational velocities (v$_{\rm br}$) of our targets by matching the synthetic line profiles to the \ion{Mg}{I} triplet at $\lambda\lambda$ 5167–5183 {\AA}, which is particularly useful for this purpose.

Regarding the computation of gravity for these stars, we modeled the lines of the \ion{Mg}{I} triplet at $\lambda\lambda$ 5167, 5172, and 5183 {\AA}, which are very sensitive to $\log$ {\itshape g} variations. First, we derived the magnesium abundance through the narrow \ion{Mg}{I} lines at $\lambda$ 5528 {\AA} (not sensitive to $\log$ {\itshape g}), and then we fitted the triplet lines by fine-tuning the $\log$ {\itshape g} value. Microturbulence velocities for these stars were estimated using the calibration $\xi = \xi(T_{eff},\log g)$ published by \citet{2004A&A...420..183A}.

The estimated atmospheric parameters are summarised in Table~\ref{tab:atm_err}.

\subsection{Chemical abundances}\label{Abundances}

Following the same procedure adopted in \citet{Trentin2023}, we choose the spectral synthesis approach to overcome problems due to the spectral line blending caused by line broadening. Synthetic spectra were generated in three steps: i) LTE atmosphere models were computed using the ATLAS9 code \citep{kurucz1993new}, using the stellar parameters in Table~\ref{tab:atm_err}; ii) stellar spectra were synthesized by using SYNTHE \citep{kurucz1981solar}; iii) finally, the synthetic spectra were convoluted for instrumental and line broadening. 

The detected spectral lines allowed us to estimate the abundances for a total of 13 different chemical elements.  For all targets we performed the following analysis: we divided the observed spectra into intervals, 25 Å or 50 Å wide, and derived the abundances in each interval by performing a $\chi^2$ minimisation of the differences between the observed and synthetic spectra. The minimisation algorithm was written in {\it IDL} language, using the {\it amoeba} routine. 

We considered several sources of uncertainties in the abundance: 
$\delta T_{\mbox{\scriptsize eff}}$, $\delta \log g$, and $\delta\xi$. According to our simulations, those errors contribute $\approx \pm$\,0.1~dex to the total error budget. Total errors were evaluated by summing in quadrature this value to the standard deviations obtained from the average abundances.

The adopted lists of spectral lines and atomic parameters are from \citet{castelli2004spectroscopic}, who have updated the original parameters of \citet{kurucz1995kurucz}. When necessary we also checked the NIST database \citep{ralchenko2019nist}.

In the atmospheres of giant stars departures from LTE calculations could be not negligible.  Recently, \citet{amarsi20} explored a wide range of non-LTE corrections for a number of chemical elements. For each of them, they computed also the difference between non-LTE and LTE abundances versus metallicities, both for dwarf and giant stars. We used their grids in order to evaluate the impact of non-LTE corrections on our results. In particular, according to the low metallicity of our targets, we neglected corrections for carbon, magnesium, silicon, calcium, and barium, while we applied a correction of 0.1~dex on our LTE sodium abundance. 

A sample of the spectra for the target stars is shown in Fig.~\ref{fig:spectra}.
The final abundances are listed in Table~\ref{tab:abundances} and ~\ref{tab:abundances_fe} (for the ease of the reader, in this table the abundances are expressed as [X/Fe] instead of [X/H]).

\begin{figure}
   \centering
   \includegraphics[width=\hsize]{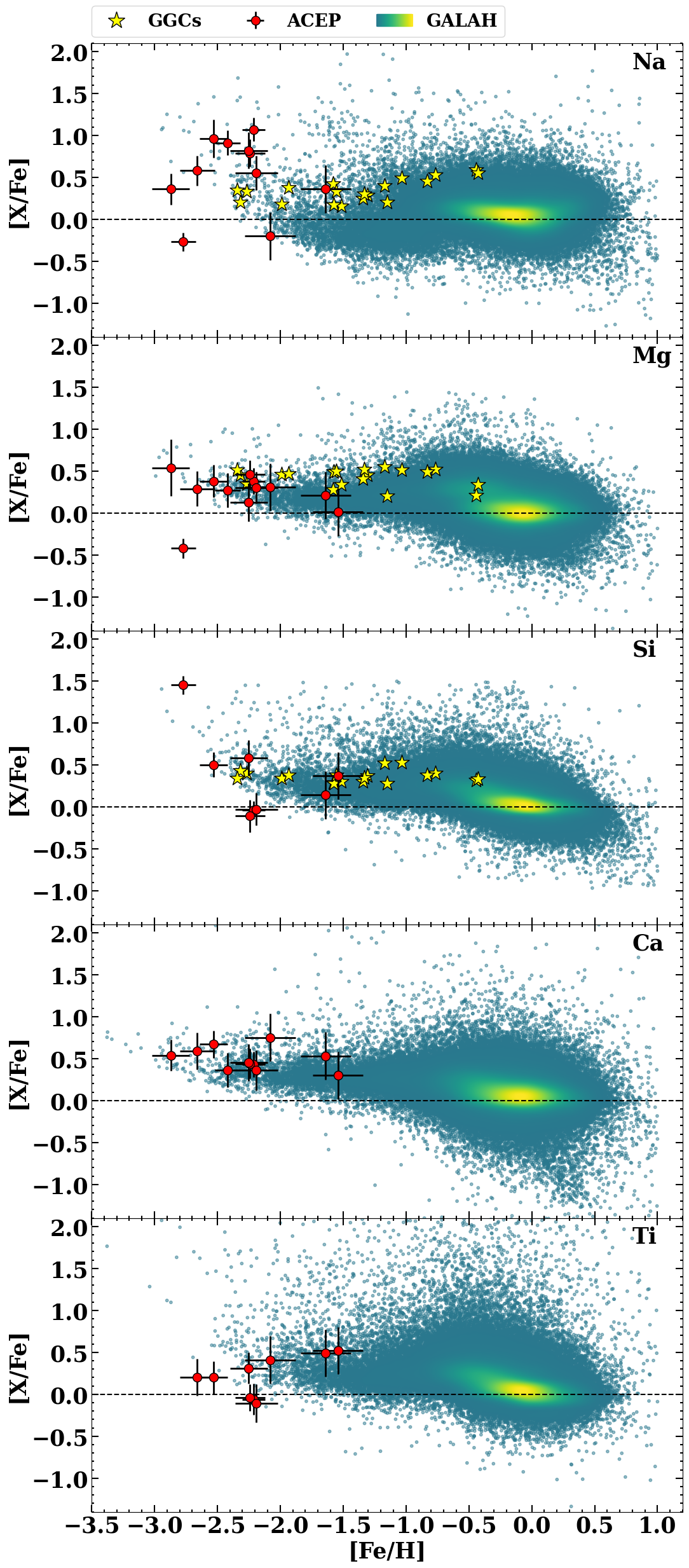}
      \caption{Abundances of Sodium, Magnesium, Silicon, Calcium and Titanium for the ACEPs investigated in this paper (solid red points with error bars) in comparison with GGCs (yellow stars) and the GALAH MW data.  
       }
         \label{fig:gcc_light}
\end{figure}

\begin{figure}
   \centering
   \includegraphics[width=\hsize]{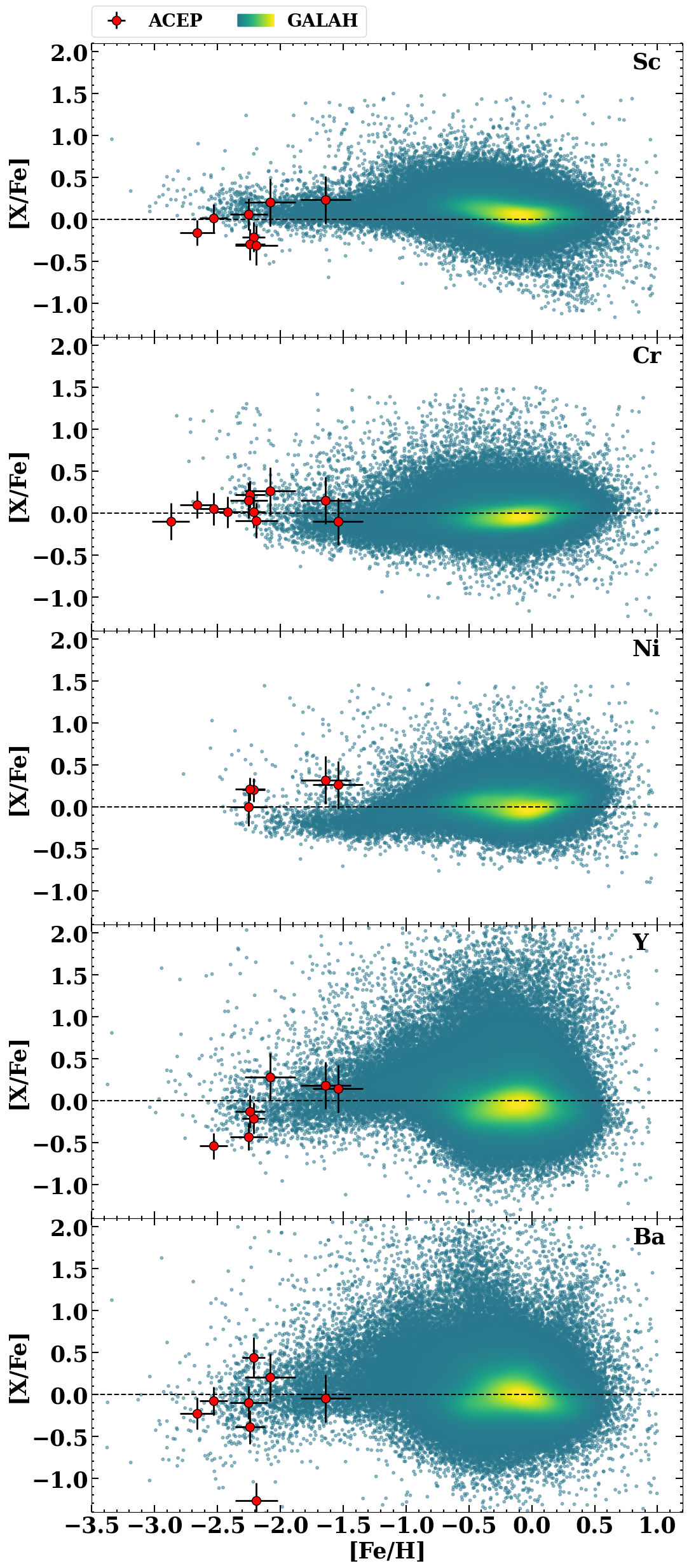}
      \caption{As in Fig.~\ref{fig:gcc_light} but for iron group and heavy elements: Sc,Cr,\,Ni,\,Y,\,Ba.  
       }
         \label{fig:gcc_heavy}
\end{figure}

\begin{figure}
   \centering
   \includegraphics[width=\hsize]{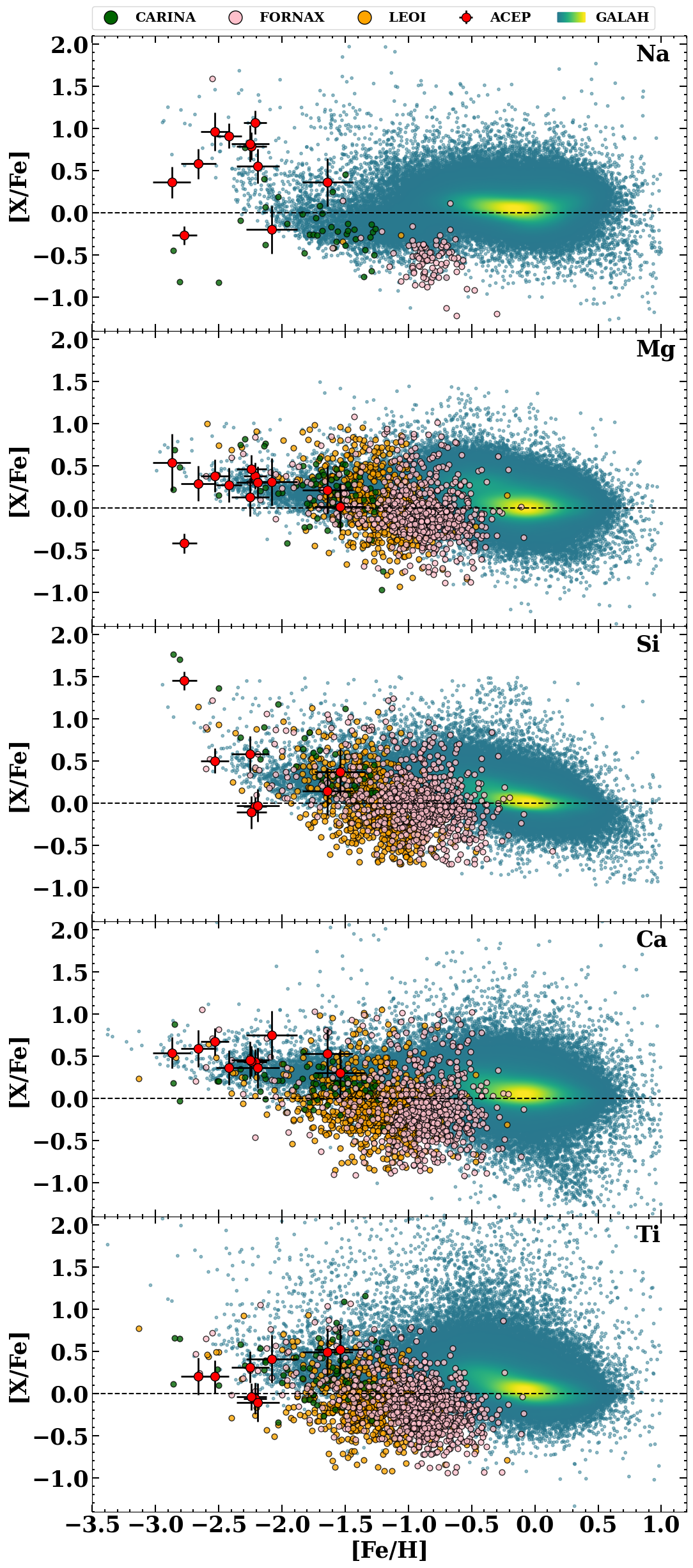}
      \caption{Same as in Fig.~\ref{fig:gcc_light} but for the comparison with three dSph hosting an extended intermediate age population, namely Carina, Fornax and Leo\,I (see labels).  
       }
         \label{fig:inter_light}
\end{figure}

\begin{figure}
   \centering
   \includegraphics[width=\hsize]{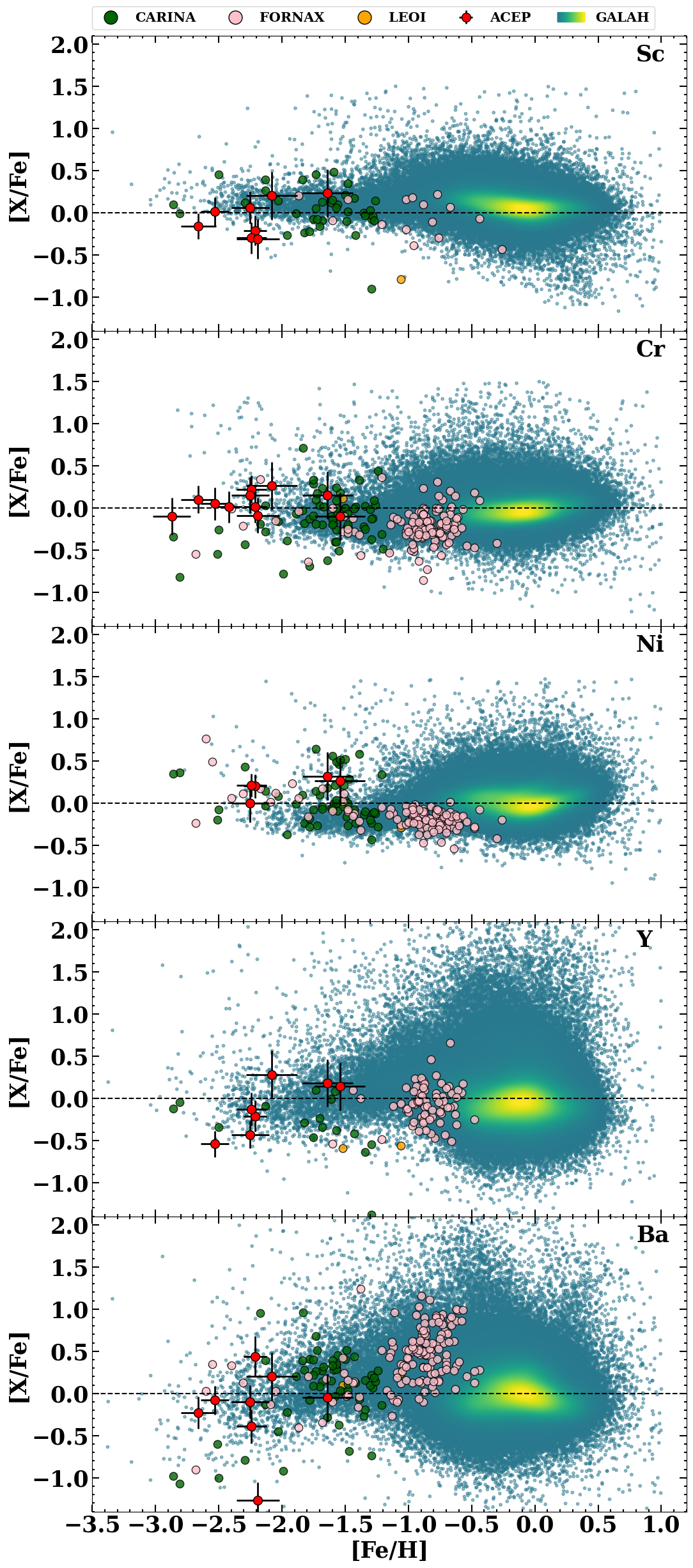}
      \caption{As in Fig.~\ref{fig:inter_light} but for the Sc,\,Cr,\,Ni,\,Y,\,Ba.  
       }
         \label{fig:inter_heavy}
\end{figure}

\begin{figure}
   \centering
   \includegraphics[width=\hsize]{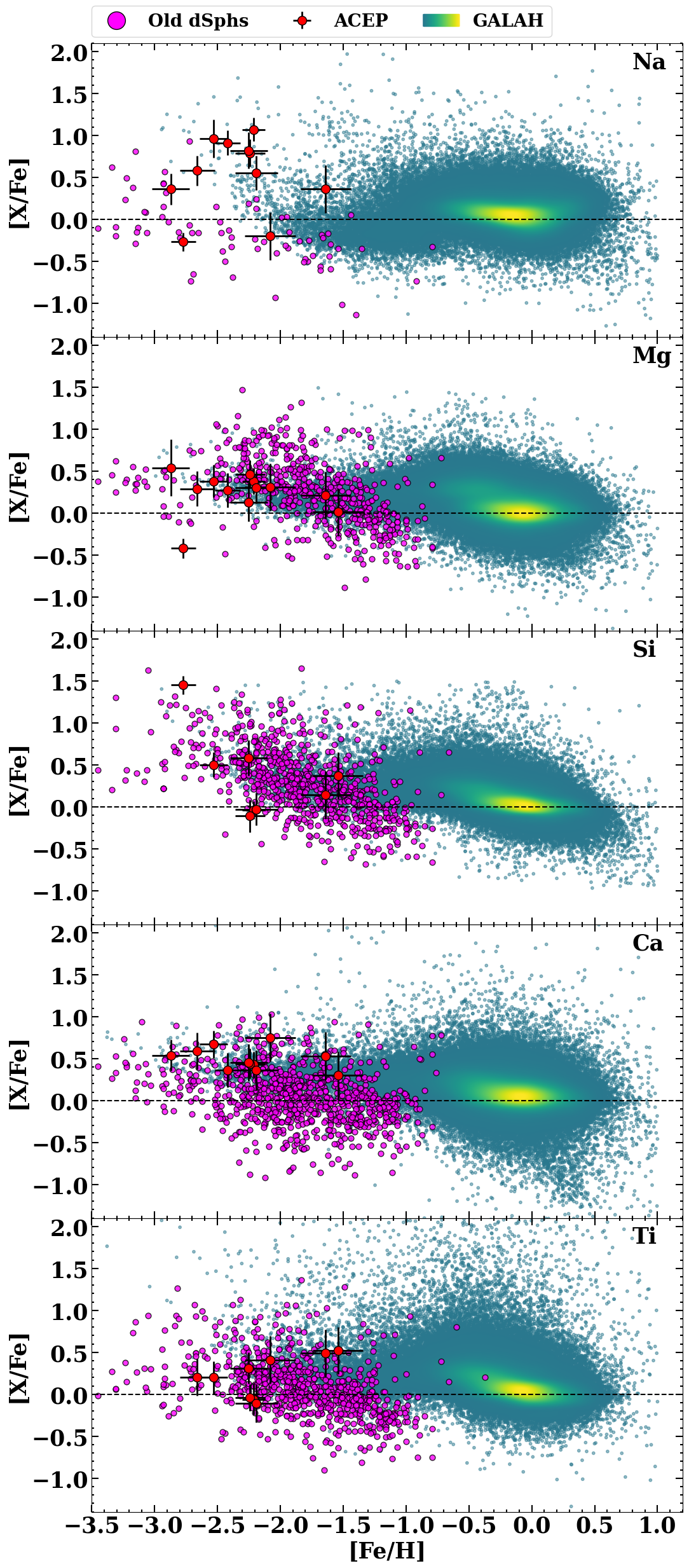}
      \caption{Same as in Fig.~\ref{fig:gcc_light} but for the comparison with several dSph hosting a purely old population. Given the high number of galaxies, for clarity reasons, we displayed all the data with the same colour (magenta).  
       }
         \label{fig:old_light}
\end{figure}

\begin{figure}
   \centering
   \includegraphics[width=\hsize]{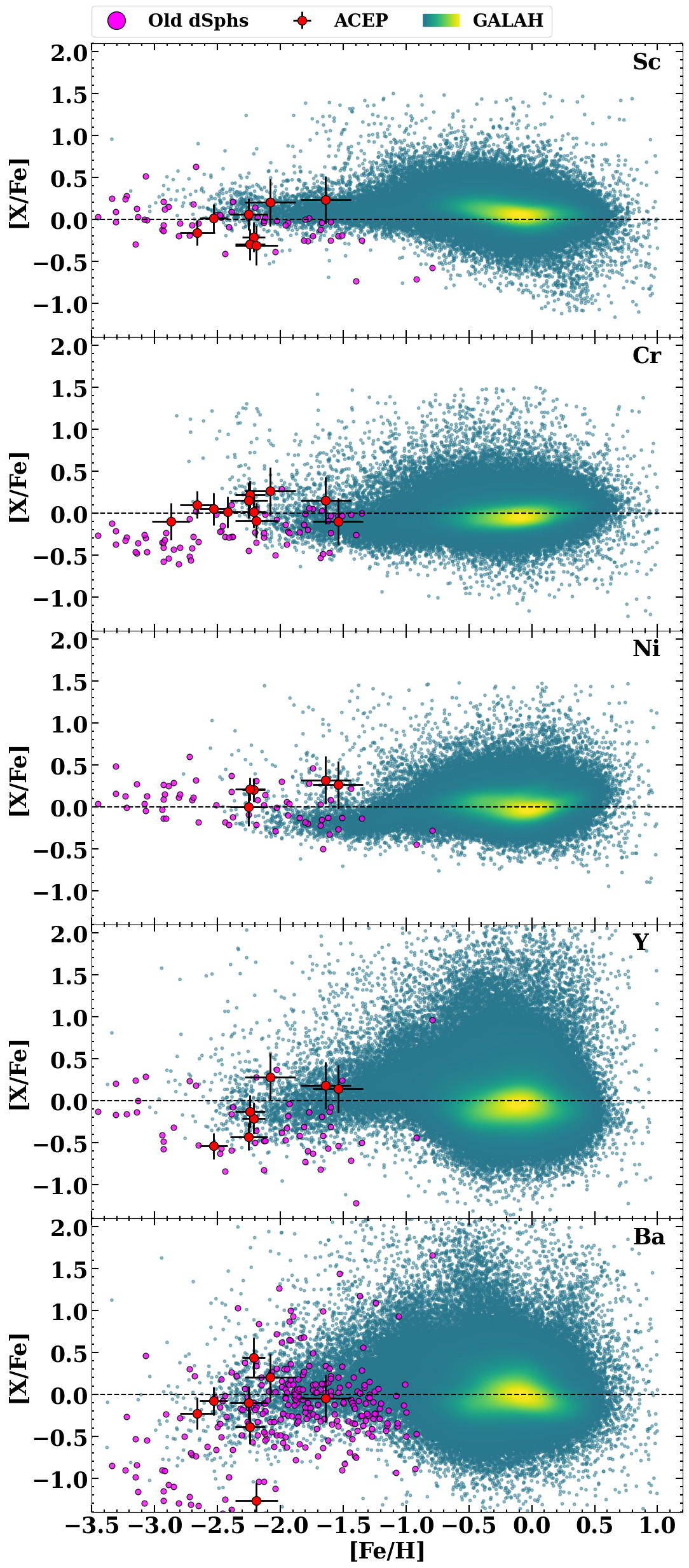}
      \caption{As in Fig.~\ref{fig:old_light} but for the Sc,\,Cr,\,Ni,\,Y,\,Ba.  
       }
         \label{fig:old_heavy}
\end{figure}

\begin{figure}
   \centering
   \includegraphics[width=\hsize]{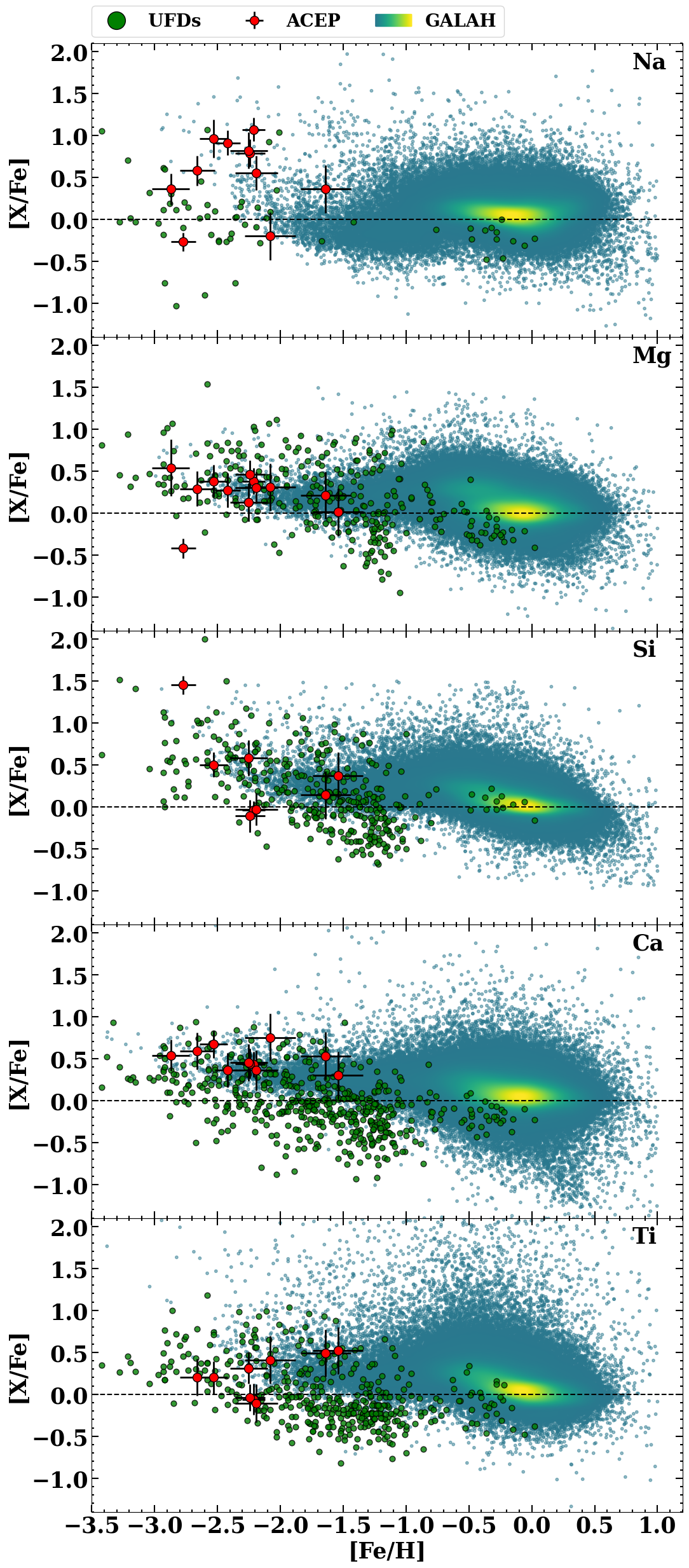}
      \caption{Same as in Fig.~\ref{fig:gcc_light} but for the comparison with several UFDs.  Given the high number of galaxies, for clarity reasons, we displayed all the data with the same colour (green).  
       }
         \label{fig:ufd_light}
\end{figure}

\begin{figure}
   \centering
   \includegraphics[width=\hsize]{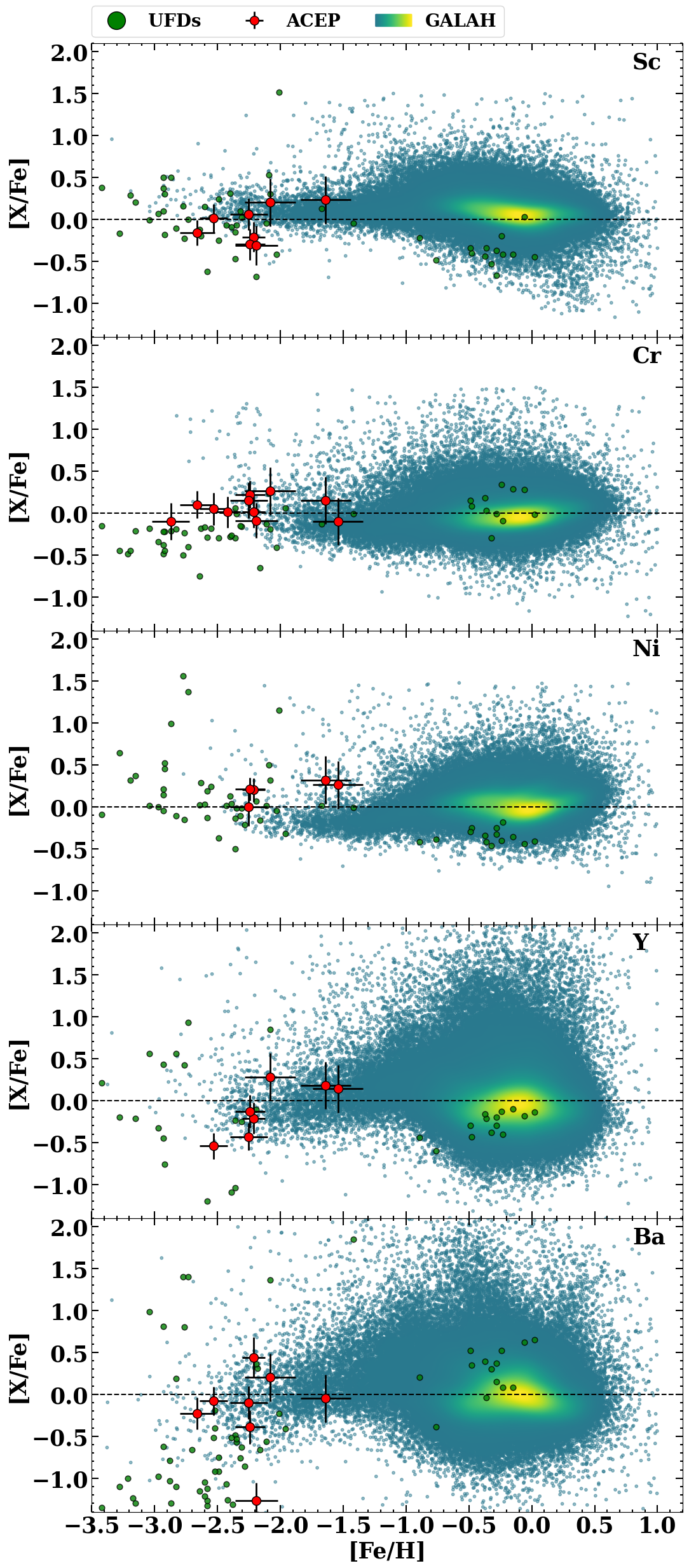}
      \caption{As in Fig.~\ref{fig:ufd_light} but for the Sc,\,Cr,\,Ni,\,Y,\,Ba.  
       }
         \label{fig:ufd_heavy}
\end{figure}

\begin{figure}
   \centering
   \includegraphics[width=\hsize]{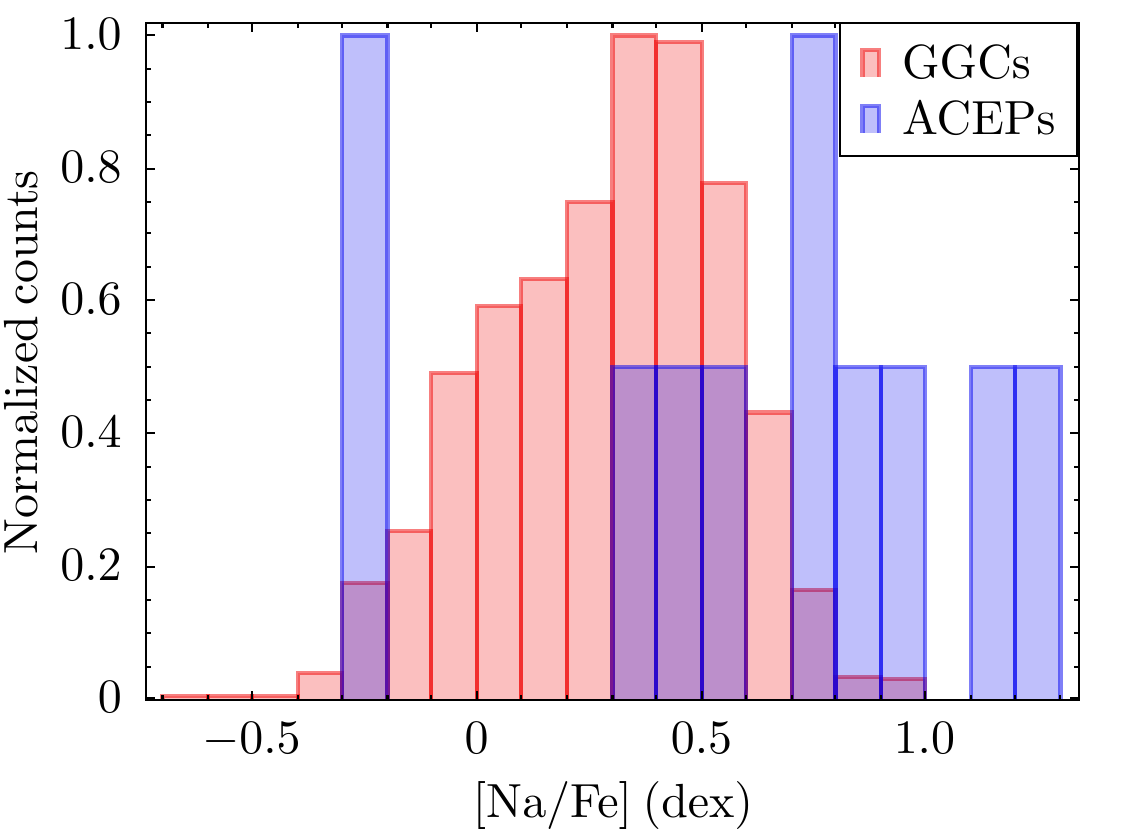}
      \caption{Histogram of individual [Na/Fe] abundances for stars belonging to 15 GGCs (red) and the investigated ACEPs (blue). The GGCs data is from 
       \citet{Carretta2009b}}
         \label{fig:histona}
\end{figure}

\section{Results}

To infer hints about the origin of the ACEPs analysed in this work, we compare their chemical abundances with those of stellar systems which could possibly resemble the place of origin of the ACEPs, namely, the MW field, GGCs, dSph and UFDs. For the MW we adopted the large sample of stars published in the DR3 of the GALactic Archaeology with HERMES (GALAH) survey \citep[][]{Buder2021}, selecting only stars with high-quality results (flag=0 for each element) and with Galactic latitude $b>|15|$ deg, to avoid disk stars which would add confusion. For the GGCs we adopted the results by \citet{Carretta2009a}, while for the dwarf satellites of the MW, we used the vast and updated database SAGA \citep[Stellar Abundances for Galactic Archaeology Database][]{Suda2008,Suda2017}.  

We divided the chemical species into two groups, i) light and $\alpha$-elements including Na (Carbon is not discussed as we have measured it only in four stars), Mg, Si, Ca and Ti; ii) iron group (including also Sc which is intermediate between $\alpha$ and iron) Cr and Ni and neutron-capture elements, Y and La. 

To discuss the properties of the investigated ACEPs, in Fig.~\ref{fig:gcc_light} and ~\ref{fig:gcc_heavy} we compare their abundances with those of the MW and GGC (when available). In a similar way, Fig.~\ref{fig:inter_light},~\ref{fig:inter_heavy}; Fig~\ref{fig:old_light},~\ref{fig:old_heavy} and Fig.~\ref{fig:ufd_light}, ~\ref{fig:ufd_heavy} show the comparison between the ACEPs abundances and those of dSphs hosting an extended intermediate-old, dSphs with a purely old stellar population and the abundances of UFDs, respectively.

In the following, we will discuss each element separately. 

\begin{itemize}

\item{\bf Sodium}
In 9 over 11 ACEPs (EK\,Del has no Na measurement) we find a significant overabundance of Sodium compared to all the comparison samples: average Halo abundances (only a few GALAH stars with similar over-abundances), Local Group dwarfs and GGCs (average values). For the GGCs, we compared ACEPs Na abundances with those of individual cluster stars. This is reported in the histogram of  Fig.~\ref{fig:histona} where we show the Na distribution of 1312 stars in 15 GGCs \citep[data from][]{Carretta2009b} in comparison with the 11 ACEPs with Na measurement. Several ACEPs have Na abundances in the region of the most extreme second-generation GGC stars or beyond. In particular, only about 4\% of GGC stars have [Na/Fe]$>$0.7 dex, while more than half of our sample shows values beyond this threshold.    
The two stars showing a small underabundance of Na are BF\,Ser and SHM2017\,J000.07389-10.22146; V716\,Oph has [Na/Fe]$\sim$0.35 dex, while all the other ACEPs have [Na/Fe]$>$0.5 dex. In particular, HE\,0114-5929, DF\,Hyi and   CRTS\,J003041.3-441620 show values of [Na/Fe] larger than 1 dex. 
We will try to explain this significant feature in the following section. 

\item{\bf Magnesium}
Similar to other $\alpha$ elements that we can measure (Ca, Si, Ti), Mg is expected to be enhanced at low iron abundance values, even if in a different way compared to the other three elements, as it is produced during the hydrostatic He burning in massive stars, and it is, therefore, less directly connected to the SN II explosion conditions \citep[e.g.][]{Tolstoy2009}. Nevertheless, Fig.~\ref{fig:gcc_light} shows that we observe Mg enhancement for all the stars, with the significant exception of SHM2017\,J000.07389-10.22146 which has [Mg/Fe]$\sim -0.4$ dex and EK\,Del, which has [Mg/Fe]$\sim 0$ dex. In general, the location of the ACEPs in the [Mg/Fe] vs [Fe/H] diagram is in very good agreement with those of the Galactic halo stars and GGCs, as well as with those of the dwarf satellites of the MW.   

\item{\bf Silicon and Titanium}
We discuss these two $\alpha$ elements together because they show similar features. 
For both elements, we could not obtain a measure in DR2\,6498717390695909376 and HE\,0114-5929. Silicon is absent also for DR2\,2976160827140900096 and BF\,Ser, while Ti is not measurable in SHM2017\,J000.07389-10.22146. 
Considering the stars for which we have a measurement, Si and Ti are enhanced, as expected, but with three notable exceptions: the ACEPs HE\,2324-1255, CRTS\,J003041.3-441620 and DF\,Hyi all show slightly negative values of both [Si/Fe] and [Ti/Fe] abundances, in contrast with the remaining 5 and 6 stars (in Si and Ti, respectively). The comparison with the MW field and GGCs (only Si) shows good agreement only with stars showing Si/Ti enhancement (see Fig.~\ref{fig:gcc_light}). The comparison with MW satellites shows a good agreement only for the ACEPs with [Fe/H]$<-2.2$ dex, especially for the UFDs. The position of the three ACEPs with low [Si/Fe] and [Ti/Fe] values seems to be more consistent with the dwarfs possessing a pure old population such as Draco or Ursa Minor, but both these galaxies show a large range of Si and Ti abundances at fixed iron. 
                                                                                         
\item{\bf Calcium}
Figure~\ref{fig:gcc_light} shows that the Calcium abundance of all the ACEPs considered here (except SHM2017\,J000.07389-10.22146 for which we could not measure this element) is enhanced and is in the range 0.3$<$[Ca/Fe]$<$0.75 dex, as expected for this pure $\alpha$ element. The comparison with the Galactic halo shows a generally very good agreement suggesting that both the ACEPs and the metal-poor halo stars were born at early times from material highly enriched by SN II ejecta \citep[SN IIs produce large amounts of $\alpha$ elements in contrast with SN Ia which mostly provide iron, see e.g.][]{Tolstoy2009}. The comparison with all the MW satellites shows that at fixed iron abundance, the ACEPs' [Ca/Fe] values are systematically larger, especially looking at the purely-old dSphs and UFDs (Figs.~\ref{fig:old_light} and ~\ref{fig:ufd_light}.

\item{\bf Scandium}
 We have measures of Sc for 7 stars, i.e. no data for DR2\,6498717390695909376, HE\,0114-5929, SHM2017\,J000.07389-10.22146 and EK\,Del. Figure~\ref{fig:gcc_heavy} shows that the [Sc/Fe] value is negative for four stars, namely DR2\,2976160827140900096, HE\,2324-1255, CRTS\,J003041.3-441620 and DF\,Hyi. Notably, the last three are the same stars that showed low [Si/Fe] and [Ti/Fe] values compared with the other ACEPs. The presence of such an underabundance of Sc for more than 50\% of our stars is surprising, given that scandium is expected to be mainly produced during core-collapse SN events, with a small contribution from AGB stars\citep[][]{Kobayashi2020}. Indeed, there are virtually no stars in the GALAH sample with a negative value of [Sc/Fe] at the relevant [Fe/H] interval. The same applies to the dSphs with an extended intermediate-age population, while some negative [Sc/Fe] values can be found among the purely old dSphs a UFDs. 
 In any case, the similarity of behaviour of the Si, Sc, and Ti for the three stars mentioned before could hint towards a common origin of these three objects, even if they lay in very different regions of the Galactic halo.

\item{\bf Chromium} 
We have measured Cr in all the stars except SHM2017\,J000.07389-10.22146. Chromium is an iron-peak element and is produced both in core-collapse and thermonuclear SNe. The measures for the ACEPs show values of [Cr/Fe] around zero within the errors, in fair agreement with the MW halo up to [Fe/H]$\sim-2.0$ dex, beyond this value there are only a few measures. Compared with the dSph and UFD data, our Cr abundances appear systematically larger. 

\item{\bf Nickel}

We have Nickel measurements only for five stars. All the measures are overabundant compared to iron except DR2\,2648605764784426624 which has [Ni/Fe]$\sim$0 dex. The bulk of the Galactic data shows negative values of [Ni/Fe] at the [Fe/H] value of interest, even if there are scattered stars with positive values. The dSphs show a dispersion of values for [Ni/Fe] but with a tendency towards negative values at the relevant metallicity range.

\item{\bf Yttrium}
Yttrium is a neutron-capture element mainly formed through s-process in AGB stars at solar metallicities, while at low metallicities a contribution from the r-process is required to explain its abundance \citep[][]{Kobayashi2020}. We have measures of Yttrium for seven stars, three and four stars showing positive and negative [Y/Fe] values, respectively (see Fig.~\ref{fig:gcc_heavy}). 
Two of the stars having negative [Y/Fe], namely DF\,Hyi and CRTS\,J003041.3-441620, are among the triad of objects with low values of Si, Ti and Sc abundances (we do not have measures for the third star HE\,2324-1255). 
Compared with the Galactic and dSphs/UFDs sample the overall distribution of the seven ACEPs in the [Y/Fe] vs [Fe/H] is in good agreement, albeit in the MW satellites the measures are scarce and rather scattered.

\item{\bf Barium}
Barium is an almost pure s-process element, but its high value at low [Fe/H] requires r-process to be at work \citep[][]{Kobayashi2020}. We have Ba abundances for eight stars as shown in Fig.~\ref{fig:gcc_heavy}. These measures show a significant degree of dispersion, in accord with the Galactic and dSphs abundances. The ACEP HE\,2324-1255 shows a significantly low value of [Ba/Fe] (this is one of the triads with low Si, Ti, Sc, and possibly Y), albeit stars with similar abundance can be found among the dSphs. 

\end{itemize}

\begin{figure}
   \centering
   \hbox{
   \includegraphics[width=9cm]{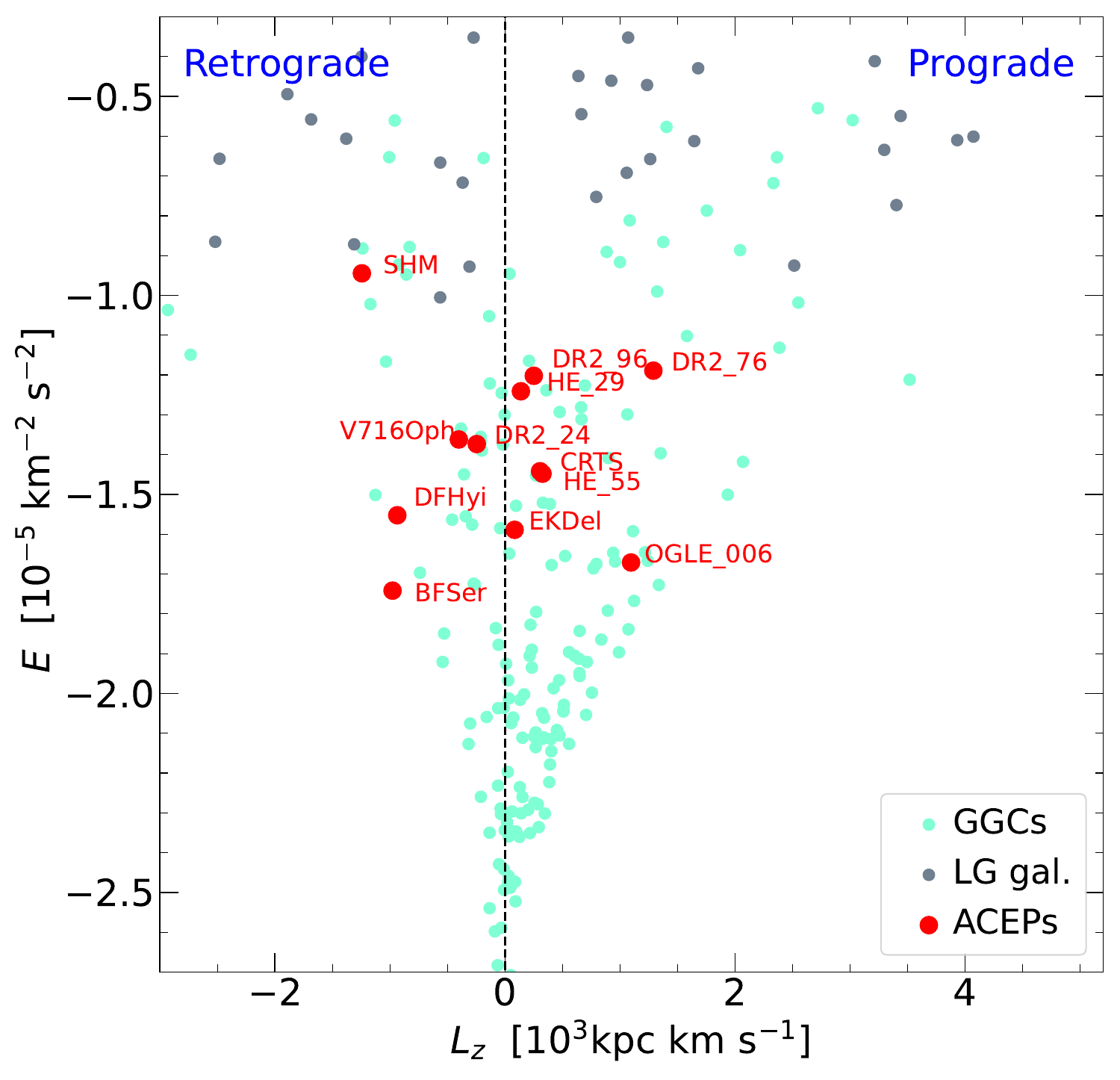}
}
      \caption{
      Angular momentum in the z direction versus total energy of the orbit of the ACEP sample studied in this paper (red-filled circles) in comparison with a sample of 166 GGCs (smaller light green filled circles) and of dwarf galaxies of the Local Group (smaller grey filled circles). \\
      Note that to avoid confusion, the long identification names of some stars were abbreviated as follows: 
CRTS=CRTS\,J003041.3-441620;  DR2\_96=DR2\,2976160827140900096; 
HE\_29=HE\,0114-5929; 
HE\_55=HE\,2324-1255; 
OGLE\_006=OGLE\,GAL-ACEP-006; 
SHM=SHM2017\,J000.07389-10.22146; 
DR2\_24=DR2\,2648605764784426624; 
DR2\_76=DR2\,6498717390695909376. 
}
         \label{fig:lz_e}
\end{figure}

\section{Discussion}

The abundance analysis presented here provides new insights into the longstanding debate on the origin of the ACEPs. In this section, we consider old and new hypotheses, forming a basis for future investigations in order to give a definitive answer to the many questions still open.


\subsection{Are the investigated ACEPs metal-poor objects?}
 
All the investigated ACEPs are metal-poor with $-2.87 \leq {\rm [Fe/H]} \leq-1.54$ dex. Albeit based on a restricted sample of ACEPs, this result represents a direct empirical confirmation of the theoretical prediction, which does not forecast the possible occurrence of ACEPs for values of [Fe/H] larger than $\sim -1.5$ dex \citep[see e.g.][]{Fiorentino2006,Monelli2022}. 
A spectroscopic investigation of the ACEPs detected in the direction of the Bulge may be a good test for such an expectation. Indeed, in this region, the intermediate-age and old stellar populations (from which the ACEP originate) can be (significantly) more metal-rich than [Fe/H]$\sim -1.5$ dex \citep[see the review by ][for the Red Clump and RR Lyrae stars]{Kunder2022}.


\subsection{Do the investigated ACEPs have an extragalactic origin?}

The MW ACEP composition resulting from the present analysis is in good agreement with that of a typical pop II star belonging to the Galactic halo, with the partial exception of sodium. Indeed, in Fig.~\ref{fig:gcc_light}, we observe a number of Galactic stars with [Fe/H$<-1.5$ dex and [Na/Fe]$>$+0.5 dex, i.e. the range where more than half of the ACEPs are placed. However, the percentage of such Galactic stars is less than 10\% of all the GALAH stars with [Fe/H$<-1.5$ dex. Therefore, even if the statistics are poor, a difference in the sodium abundance between the ACEPs and the metal-poor population of the Galactic Halo appears to be present. We shall come back to this point in Sect.~\ref{sect:origin_of_na}. \\    
Overall, the $\alpha$-elements, Mg and Ca, in particular, are over-abundant compared to iron. This is a well-known feature of pop II stars, which are believed to be formed from primordial gas mostly enriched with yields of core-collapse supernovae. Similarly, the heavier elements show a general agreement with the pop II Galactic sample. Worth to note, three stars, HE\,2324-1255, CRTS\,J003041.3-441620 and DF\,Hyi, 
show low [Si/Fe], but quite large [Mg/Fe] and [Ca/Fe], while SHM2017\,J000.07389-10.22146 shows a negative [Mg/Fe] (-0.42 dex), but a quite large silicon over-abundance ([Si/Fe]=+1.45 dex). 
At odds with all the other ACEPs in our sample, SHM2017\,J000.07389-10.22146 is also sodium depleted ([Na/Fe]=-0.27 dex).  One possibility to explain these anomalous compositions is to hypothesise that they are the consequence of local pollution of the gas from which these stars were born. Note that the scatter of the $\alpha$-element abundances is generally small in halo stars. For instance, in the majority of the stars with [Fe/H]<-1.5 dex, the magnesium overabundance ([Mg/Fe]) rages between $+0.1$ dex and $+0.5$ dex, with an average [Mg/Fe]$\sim 0.3$ dex   \citep[see e.g.][]{sneden_2008}. However, for a few halo stars significantly different [$\alpha$/Fe] values have been found \citep[see e.g.][]{ryan_1996,ivans_2003}. This occurrence represents evidence of a certain degree of inhomogeneity in the early Galaxy. 
\\
On the other hand, the comparison of the MW ACEP composition with that of the stellar populations harboured in dSph and UFD galaxies reveals more differences than similarities.
Apart from the sodium, discussed below, the main differences are: i) Figures~\ref{fig:inter_light},~\ref{fig:old_light} and ~\ref{fig:ufd_light} show that the distribution of the [Ca/Fe] abundances in the MW satellite appears systematically lower than that of ACEPs, albeit not completely incompatible in the case of the old dSph and UFDs; ii) similar to Ca, Fig.~\ref{fig:inter_heavy},~\ref{fig:old_heavy}, and ~\ref{fig:ufd_heavy} show that the chromium (and possibly nickel) in ACEPs is systematically more abundant than in the MW satellites. 

The most noticeable difference is in the abundance of sodium. Figures~\ref{fig:inter_light},~\ref{fig:old_light} and \ref{fig:ufd_light} show that the MW satellites do not have stars with high [Na/Fe] values at the metallicity of ACEPs. However, two objects, SHM2017\,J000.07389-10.22146 and BF\,Ser, display negative [Na/Fe] values, while V716\,Oph has [Na/Fe]$\sim$0.35 dex and are compatible with the abundances of old dSphs and UFDs. To investigate the possibility that at least these stars can originate from a merging of one of these objects, we display in the left panel of Fig.~\ref{fig:lz_e} the angular momentum in the $z$ direction $L_Z$ vs the energy of the orbit for the investigated ACEPs in comparison with those of GGCs which, in large part are thought to have been accreated during ancient mergers \citep[e.g.][]{Massari2019,Kruijssen2020}. The details concerning the calculation of these parameters are reported in Appendix~\ref{AppendixC}. The location of the three ACEPs with negative or mildly positive [Na/Fe] values is on the retrograde side of the plot (negative values of $L_Z$), a feature that can reveal an extragalactic origin. However, also DF\,Hyi and DR2\,2648605764784426624, which show high values of [Na/Fe] not present in MW satellites have retrograde motion, therefore it is difficult to draw firm conclusions about the extragalactic origin of the aforementioned three stars, also in light of the difference in calcium an chromium. An exception could be SHM2017\,J000.07389-10.22146, which displays not only a large retrograde motion but also high energy, typical of dwarf galaxy satellites of the MW, as shown in the right panel of Fig.~\ref{fig:lz_e}. Considering also its other chemical anomalies, SHM2017\,J000.07389-10.22146 is the most likely candidate among our sample of ACEPs to have an extragalactic origin.     
\\
In conclusion, a scenario in which the ACEPs found in our Galaxy were assembled in metal-poor extragalactic environments and, later on, they were captured by the gravitational potential well of the Milky Way, appears disfavored by the observed abundance pattern, with the noticeable exception of SHM2017\,J000.07389-10.22146.

\subsection{Are the investigated ACEPs single stars?}

Based on their composition and kinematics, the MW ACEPs in our stellar sample appear as typical Galactic halo stars. On the other hand, their luminosity, colour and pulsational properties, all indicate that the present mass of these core-He burning stars ranges between 1.3-2.5 $M_{\odot}$, which corresponds to single-star ages between 1 and 6 Gyr. Intermediate-age stellar populations like this are commonly found in dwarf galaxies, but halo stars are definitely older. There are, indeed, multiple pieces of evidence confirming that the star formation in the Galactic halo ceased no later than 10 Gyr ago
 \citep[see e.g.][ and references therein]{Haywood2016}.

On the other hand, the chemical pattern discussed in previous sections leads us to exclude an extra-galactic origin for the MW ACEPs. 
Therefore, it appears that these stars must be old, with ages of 10 Gyrs or larger, but their present-day mass should be larger than 1.2 $M_\odot$. The only known scenarios explaining old and massive core-He burning stars are those involving binary systems, originally made by two low-mass companions whose evolutionary lifetime is comparable or longer than the halo age \citep{Renzini1977}. 
According to these scenarios, after a sufficiently large portion of their single-star lifetime, the two components of the binary system interact, through a mass transfer process (Roche-lobe overflow) and, eventually, a coalescence. Most likely, the binary interaction occurred when the two stars were both on the main sequence. In this case, a BS would form. Then, after the central-H exhaustion, the BS evolves becoming a red giant and, later on, a core-He burning star. If the mass of this core-He burning star is in the range between 1.2 and 2.1 $M_{\odot}$, it will become an ACEP pulsator.
Alternatively, the binary interaction occurred after the primary evolved out of the main sequence when the star became a red giant star. In this case, the final outcome may be very different depending on the initial orbital parameters and if the mass transfer is conservative (total mass and total angular momentum are conserved) or not. In any case, since the donor has an extended convective envelope, its radius increases as its mass is eroded. In addition, until the mass ratio $M1/M2>1$, the two components of the binary get closer and, in turn, the mass transfer rate is expected to be quite fast at the beginning. In these conditions, the amount of mass that the secondary star may actually accept is rather uncertain. Later on, when the mass ratio becomes lower than 1, the separation increases and the mass transfer rate becomes slower. In any case, the mass transfer should continue until the whole envelope of the giant star is lost or the RGB tip is reached. In practice, the masses of the two components and the separation after this mass transfer episode depend on a complex phenomenology.  Then, if the final mass of the secondary is larger than 1.2 $M_{\odot}$ and no further mass transfers occur, it may eventually become an ACEP. \citet{Renzini1977} found that there exists a fairly large range of initial separations for which this scenario may actually result in the formation of an ACEP on a timescale of the order of 10 Gyr. However, they assumed a conservative mass transfer. 
 Note that in both the binary scenarios described above, main-sequence star merger or RGB mass transfer, the Galactic ACEPs would be descendants of BSs.
\\
In summary, the present data suggest that the Galactic ACEPs investigated in this paper should have originated in a binary star.

\subsection{Which stars may synthesise sodium and which abundances may correlate with sodium enhancement?}


Sodium is produced by proton captures on $^{22}$Ne in the H-burning shell of RGB, AGB and massive-giant stars. When a deep-mixing process sets on, additional $^{22}$Ne is moved down to the layers where the H-burning takes place, while freshly synthesised $^{23}$Na is dredged up to the stellar surface. Extant nucleosynthesis models show that in order to obtain such a sodium enhancement, temperatures as large as 20-30 MK should be attained by this external circulation (see next section). This is known to occur in massive AGB stars (M$>3-4$ M$_{\odot}$), in which the convective envelope penetrates inward down to the hottest layers of the H-burning shell. This process is called Hot Bottom Burning (HBB). 
In massive stars (M$>20-25$ M$_{\odot}$), the required deep mixing may be activated in case of fast rotation \citep[e.g.][]{Decressin2007}. In all other cases, such as in RGB stars, also other instabilities could induce mixing below the convective envelope, such as gravity waves generated at the convective border \citep[][]{Denissenkov2003}, magnetic buoyancy \citep[][]{Busso2007} or thermohaline circulation \citep[][]{Charbonnel2007}. 
In all these cases, the Na enhancement should be associated with a significant N enhancement, while O should be depleted.     
Unfortunately, we do not have O and N lines in our spectra, but the search for these correlations/anti-correlations may be the goal of future investigations. 
Na can be also produced by neutron capture on  $^{22}$Ne, followed by the $\beta$ decay of the radioactive $^{23}$Ne. In a low-mass AGB star, a $^{13}$C pocket forms after each third dredge-up episode, and, later on, neutrons are released by the $^{13}$C$(\alpha,n)^{16}$O reaction \citep[][]{Straniero2006,Cristallo2009}.  
This is the same environment where the bulk production of the s-process main component occurs in nature. In that case, the third dredge-up is also responsible for the atmospheric sodium enrichment, which should be associated with strong enhancements of both C and s-process elements. For a few ACEPs, we have observed Y and Ba and these measurements seem to exclude that the Na enhancement is a consequence of a low-mass AGB pollution. Indeed, the measured abundances of these heavy elements are those typically found in halo stars, in which they originate from the r-process, rather than from the s-process \citep[][]{sneden_2008}.   

\begin{figure}
   \centering
   \includegraphics[width=\hsize]{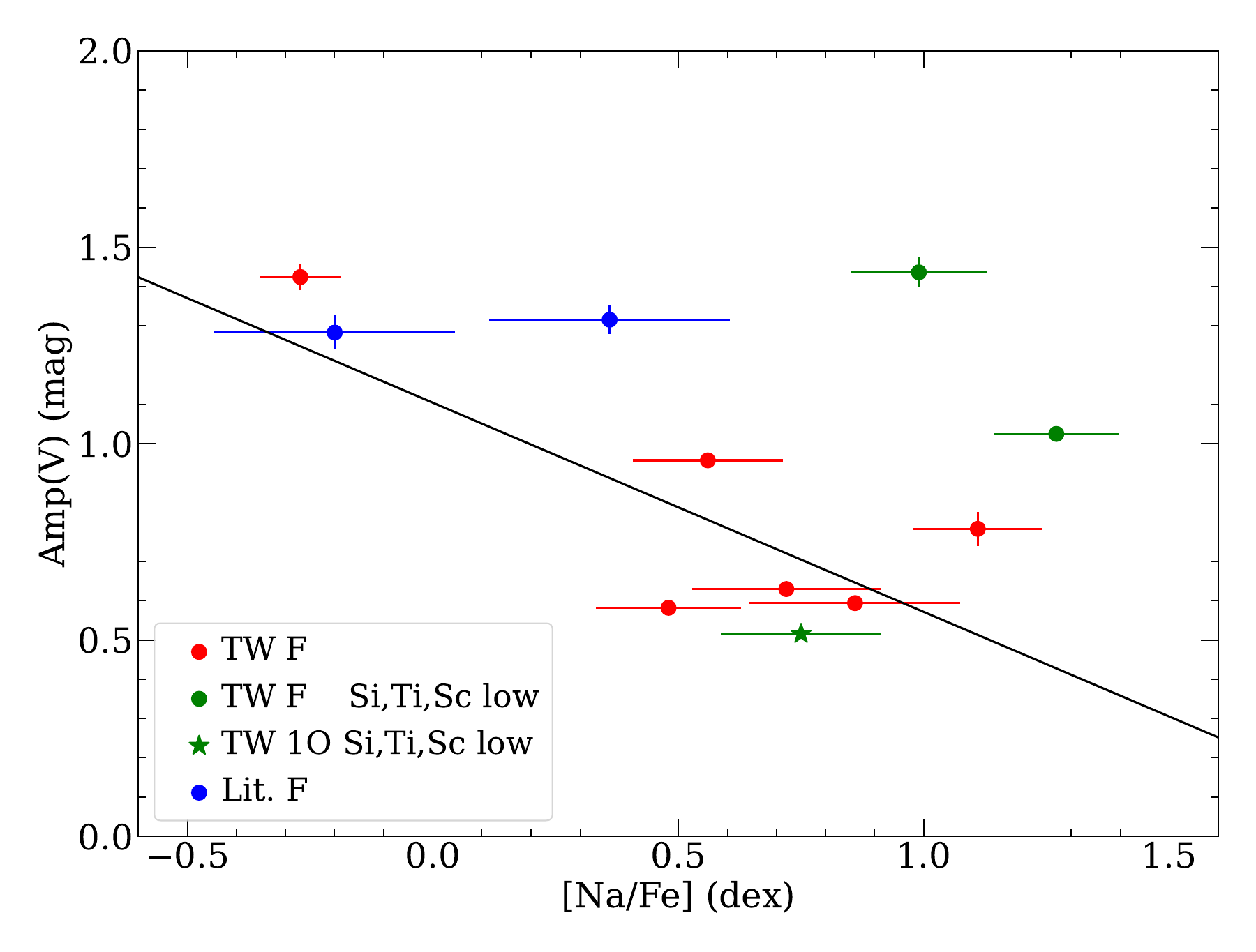}
      \caption{Visual amplitude of the observed ACEPs as a function of the [Na/Fe] abundance. The solid line shows a linear regression to the data, not including the three stars with low values of the [Si/Fe], [Ti/Fe] and [Sc/Fe] abundances (green symbols, see labels). In the labels, TW and Lit. indicate the nine stars studied spectroscopically in this paper and the three with spectroscopy from the literature, respectively.    
       }
         \label{fig:amp_na}
\end{figure}

Let us finally note that both the HBB, active in massive AGB stars, and the rotation-induced mixing, eventually active in fast-rotating massive stars, should also produce a substantial He enhancement. A certain He enhancement is also expected in case of activation of a deep-mixing process in RGB and low-mass AGB stars. The He abundance cannot be easily measured in ACEP stars. However, according to \citet{Kovtyukh2018}, a helium over-abundance should affect the light curves of these pulsators. Indeed, the theory predicts that for a variety of pulsating variables,  higher He implies smaller amplitude and more sinusoidal (i.e. less-asymmetric) light curves \citep[][]{Marconi2016}. Therefore, we searched for a correlation between the abundance of Na and the amplitude in $G$ mag of the MW ACEPs.
This test is shown in Fig.~\ref{fig:amp_na}. The expected correlation is clearly recognised (Correlation coefficient=-0.79), especially if we do not consider the two ACEP\_F CRTS\,J003041.3-441620 and DF\,Hyi, which show low Si, Ti and Sc abundances (the third one,  HE\,2324-1255 is a 1O pulsator and has low amplitude for this reason), thus possibly showing an additional anomaly compared with the other ACEPs. 
In summary, there is more than one channel to synthesise Sodium in AGB or massive stars. In any case, an overabundance of Na should come along with an overabundance of He. Pulsational arguments suggest that in the investigated stars there is a correlation between the [Na/Fe] value and the He abundance.  



\subsection{What is the origin of the [Na/Fe] overabundance?}

\label{sect:origin_of_na}

As already remarked, the most notable result of our analysis is the large [Na/Fe] values we found in 9 out of 11 stars in our sample for which the Na values were measurable. A large overabundance of sodium, associated with nitrogen and, possibly, helium enhancements, and oxygen depletion, 
is typical of some Globular Cluster stars. According to the most popular scenario, the Na overabundance is the result of a self-pollution. The polluters would have been first-generation AGB stars or fast-rotating massive stars, while the second generation of stars, as formed from the gas lost by the polluters, are those showing the mentioned chemical peculiarities 
\citep[see e.g.][and references therein]{Gratton2012}. On the contrary, the large majority of the field halo stars show scaled solar sodium. Worth noting is the discovery of a possible Na-O anticorrelation recently found in a sample of type II Cepheids (BL Her pulsators) \citep{Kovtyukh2018}.    

\begin{figure}
   \centering
   \includegraphics[width=\hsize]{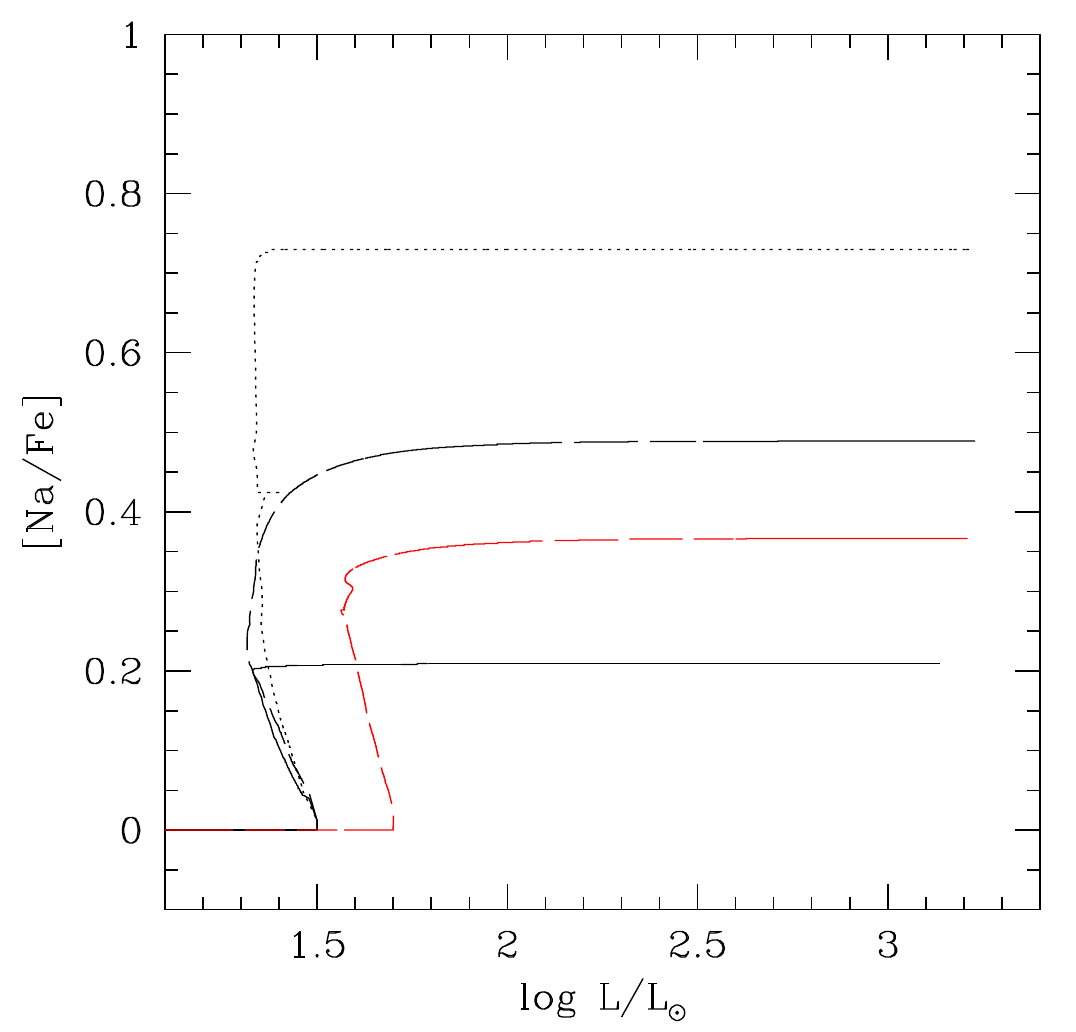}
      \caption{Sodium overabundances in RGB stellar models undergoing a rotation-induced deep-mixing process. Black lines refer to 1.2 $M_\odot$ models, while the red one to a 1.5 $M_\odot$. Solid, long-dashed and dotted lines represent cases in which the deep-mixing attains a maximum temperature $T_{max}=19$, 23 and 27 MK, respectively.   
       }
         \label{fig_RGB_na}
\end{figure}

In the previous sections, based on the observed chemical pattern we have excluded an extragalactic origin for the MW ACEPs. Hence, if they are field stars of the MW halo, how can we explain the Na overabundance?
There are multiple possibilities:
\begin{itemize}
\item 
{\bf The MW ACEP progenitors were Na-rich GC stars.}
Several problems affect this hypothesis. 
First of all,  ACEPs are very rare in GCs, there are only two confirmed objects (see Introduction). In addition, as previously discussed, MW ACEPs are more likely the descendants of interactive binaries and the estimated ACEP masses are substantially larger than that of the other cluster stars so they are expected to be more centrally concentrated in a GC and more gravitationally bound. Finally, the number of second-generation stars is comparable to those of the first-generation. Therefore we should observe a comparable number of ACEPs with and without the sodium enhancement, which is in contrast with our results.

In a recent paper, \citet{Belokurov2023} found that about 4\% of the Galactic stars with [Fe/H]$<-1.5$ dex show high [N/O] abundances, which would also imply high [Na/Fe] values. Indeed, according to the Authors, these objects originated from second-generation stars produced at early times in now completely disrupted GCs. Of course, these stars are the perfect progenitors for the ACEPs we have analysed in this work, but statistical considerations suggest that they can only account for one object maximum, as we observe high [Na/Fe] in the 9 over 11 ACEPs. 
In conclusion, the hypothesis that MW ACEPs were born in GCs appears quite unlikely. 

\item 
{\bf Na enrichment as a consequence of a mass transfer from a massive AGB or a fast rotating massive star companion in a binary system}.
As already mentioned, the wind of massive AGB stars experiencing HBB or that of fast-rotating massive stars, in principle may be the source of the required Na enrichment. 
However, the involved timescales are too short. 
Indeed, to transfer a sufficient amount of Na, the donor must be a rather massive object ($\geq 4  M_{\odot}$), In the most favourable case, the accretion occurred just $\sim 0.1$ Gyr since the binary formation. Even considering a final accretor mass as low as 1.2 $M_\odot$ (the minimum mass for an ACEP), the total lifetime is in any case no more than 6 Gyr,  definitely smaller than the halo age.
In addition, when a massive AGB experiences a Roche-lobe overflow, a common envelope likely takes place so that most of the envelope mass would be lost. This inconvenience would be avoided in case of wind accretion.  

\item 
{\bf Na enrichment as a result of an intrinsic nucleosynthesis}:  
Present stellar models show that Na is produced in the H-burning shell of RGB stars, at a temperature between 20 and 30 MK. The temperature at the base of the convective envelope is much smaller so that, in order to see a sodium enhancement at the stellar surface, a further deep-mixing process penetrating the external tail of the H-burning shell is needed. A fast rotation may provide the engine for such a process. Indeed, 
if MW ACEPs originate from low-mass interacting binaries, main-sequence stars mergers or RGB mass transfer (see previous section), 
an efficient angular momentum transfer would cause a significant 
gain in the rotational speed of the progenitor. After this foundation event, a fast-rotating BS would form that, later, evolves into a fast-rotating red giant. Then, meridional circulations or other rotational-induced instabilities, such as
those due to shear between a fast-rotating envelope and 
a slow-rotating core, may activate a mixing process in the region between 
the inner border of the convective envelope and the H-burning shell. To quantify the resulting sodium overabundance we have computed some test models of RGB stars, with mass 1.2 and 1.5 $M_\odot$ and typical composition of metal-poor halo stars, namely: 
[Fe/H]=$-$2.1 dex, [$\alpha$/Fe]=0.4 dex and initial He mass fraction Y=0.25. These models have been computed by means of the FuNS, a stellar evolution code widely used to model stars of any mass and composition, with or without rotation, whose latest version is described in \citet{straniero_2019} and \citet{straniero_2020}.
In the present models, a deep-mixing below the convective envelope is switched on after the star attains the RGB bump luminosity. 
Before this point, the deep-mixing is prevented by the presence of the sharp molecular-weight gradient as left by the first dredge up. The RGB bump occurs when the H-burning shell reaches this layer and the $\mu$-gradient is smoothed out. The resulting Na overabundance is illustrated in Fig.~\ref{fig_RGB_na}.

Here, the deep-mixing process depends on two free parameters, namely: the maximum temperature attained at the deepest mixed layer and the average mixing velocity. The resulting sodium enhancement mainly depends on the former, while a variation of the latter has a weaker effect.

Hints in favour of this intrinsic-nucleosynthesis scenario are provided by some evidence that a large fraction of BSs are fast-rotating stars with spin rates generally above 50 km/s and up to 150 km/s \citep[e.g.][]{shara1997, lovisi_2010, mucciarelli_2014, sills_2016}. More interesting is the recent finding by \citet{ferraro_2023} that the frequency of fast-rotating BSs is higher in loose globular clusters, those where the rate of lost stars should be also higher.   

Summarising, the binary origin may simultaneously explain the average ACEP composition, which is that typical of Galactic halo stars, the anomalous Na overabundances, their relatively large masses and their long lifetime.  
 
\end{itemize}

\section{Conclusions}

We presented the first high-resolution spectroscopic analysis for a sample of 9 ACEPs belonging to the Galactic halo. The analysis of these spectra allowed us to derive the abundances of 12 elements, including C, Na, Mg, Si, Ca, Sc, Ti, Cr, Fe, Ni, Y, and Ba.      
We complemented these data with literature abundances from high-resolution spectroscopy for an additional three ACEPs which were previously incorrectly classified as type II Cepheids, bringing the investigated objects to a total of 12.  
We studied the chemical properties of the ACEPs in comparison with Galactic field stars, GGCs, dSph and UFD galaxies satellites of the MW. 
The main results of this work are:

\begin{itemize}
    \item
    All the investigated ACEPs are metal-poor, with [Fe/H]$<-1.5$ dex, thus confirming the theoretical predictions asserting that a star can enter the instability strip to become an ACEP only if it has [Fe/H] smaller than about $-$1.5 dex.

    \item 
    The abundance ratios [X/Fe] vs [Fe/H] diagrams for the different elements show that at fixed [Fe/H], the position of the ACEPs is generally consistent with those of the Galactic halo field stars as measured by the GALAH survey, with the exception of the Sodium abundance, which is overabundant in all the ACEPs but two, in close similarity with second-generation stars in the GGCs. According to recent results, a fraction of the Galactic field halo stars can derive from such kinds of stars, but from statistical considerations, they cannot explain the presence of sodium overabundance for the 80\% of the ACEPs investigated in this work.

    \item 
    The same comparison with dSphs and UFDs reveals more differences than similarities, not only for what concerns Sodium but also for other $\alpha$ elements, such as Calcium and Chromium. In general, a scenario in which the ACEPs found in our Galaxy were assembled in metal-poor extragalactic environments and, later on, they were captured by the gravitational potential well of the Milky Way, appears disfavored by the observed abundance pattern, with the noticeable exception of the star SHM2017\,J000.07389-10.22146, which has negative [Na/Fe] value and whose orbit shows large retrograde motion and has an energy comparable with that typical of dwarf galaxies satellite of the MW.   

    \item 
    To explain the global characteristics of the investigated ACEPs, and, in particular their supposed old age and peculiar chemical abundances, the most likely explanation is that they are binary stars.
    
    \item
    We explore several possibilities to explain the anomalously large abundance of Sodium in the ACEPs' atmospheres. At the moment, the best explanation is the evolution of low-mass stars in a binary system with either mass transfer or merging which results in a rotational speed-up of the star due to the conservation of angular momentum, which in turn can trigger rotational mixing, able to bring the products of internal nuclear burning to the surface. However, detailed modelling is needed to confirm this hypothesis. 
\end{itemize}

The findings reported in this work must be corroborated by the observation of a larger sample of ACEPs by means of high-resolution spectroscopy. To this end, the advent of large spectroscopic surveys such as those foreseen with the WEAVE (WHT Enhanced Area Velocity Explorer)\footnote{https://www.ing.iac.es/astronomy/instruments/weave/weaveinst.html} and 4MOST (4-metre Multi-Object Spectroscopic Telescope)\footnote{https://www.eso.org/sci/facilities/develop/instruments/4MOST.html} 
will certainly provide us with the statistics needed to draw more firm conclusions about the origin of the ACEP pulsators in our Galaxy.  

\begin{acknowledgements}
We thank our anonymous Referee for their helpful comments which helped us to improve the paper. VR wishes to thank Zdenek Prudil for his help with the use of the {\tt Galpy} package. 
This work was partially supported by the INAF-GTO program 2023: "C-MetaLL - Cepheid metallicity in the Leavitt law" (P.I. V. Ripepi).
Part of this work was supported by the German 
\emph{Deut\-sche For\-schungs\-ge\-mein\-schaft, DFG\/} project
number Ts~17/2--1.\\
This research has made use of the SIMBAD database, operated at CDS, Strasbourg, France.\\

This work has made use of data from the European Space Agency (ESA) mission
{\it Gaia} (\url{https://www.cosmos.esa.int/gaia}), processed by the {\it Gaia}
Data Processing and Analysis Consortium (DPAC,
\url{https://www.cosmos.esa.int/web/gaia/dpac/consortium}). Funding for the DPAC
has been provided by national institutions, in particular, the institutions
participating in the {\it Gaia} Multilateral Agreement.\\

M.M. acknowledges financial support from the Spanish Ministry of Science and Innovation (MICINN) through the Spanish State Research Agency, under Severo Ochoa Programe 2020-2023 (CEX2019-000920-S), and from the Agencia Estatal de Investigaci\'on del Ministerio de Ciencia e Innovaci\'on (MCINN/AEI) under the grant "RR Lyrae stars, a lighthouse to distant galaxies and early galaxy evolution" and the European Regional Development Fund (ERDF) with reference PID2021-127042OB-I00

\end{acknowledgements}

%
   \bibliographystyle{aa} 
   \bibliography{myBib} 
%

\begin{appendix} 

\section{\gaia\ light curves for the investigated ACEPs}

\begin{figure*}[h]
  \centering
\vbox{
\hbox{
\includegraphics[width=0.285\textwidth]{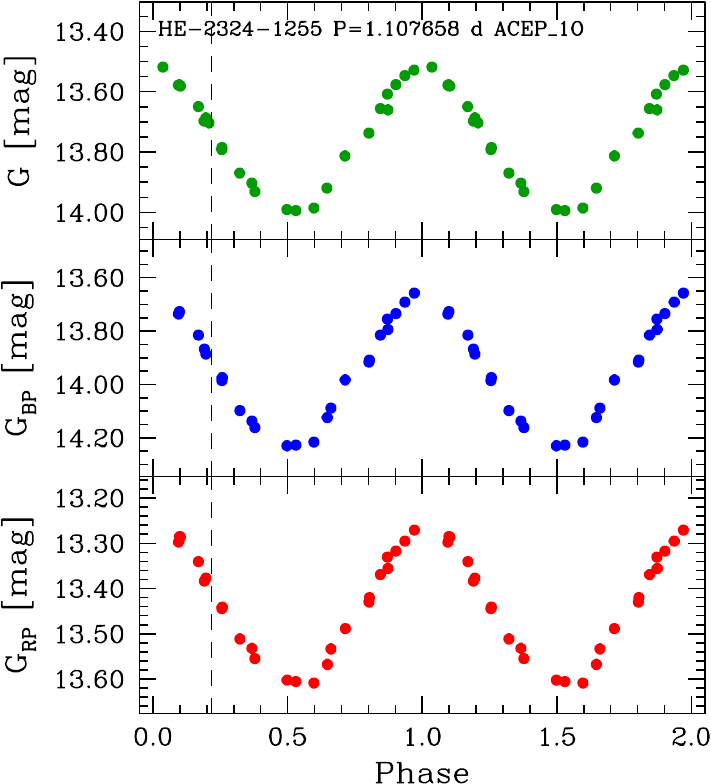}
\includegraphics[width=0.285\textwidth]{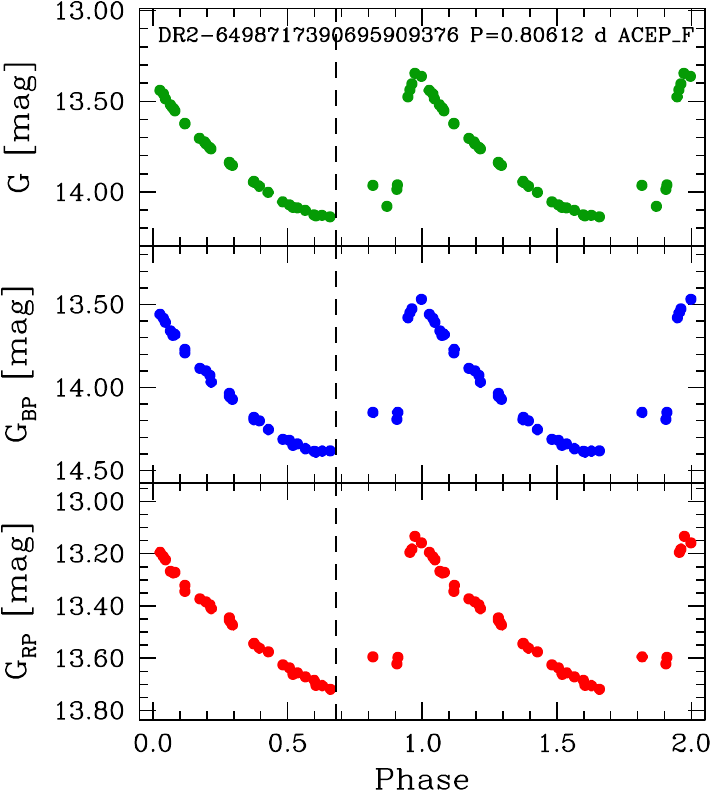}
\includegraphics[width=0.285\textwidth]{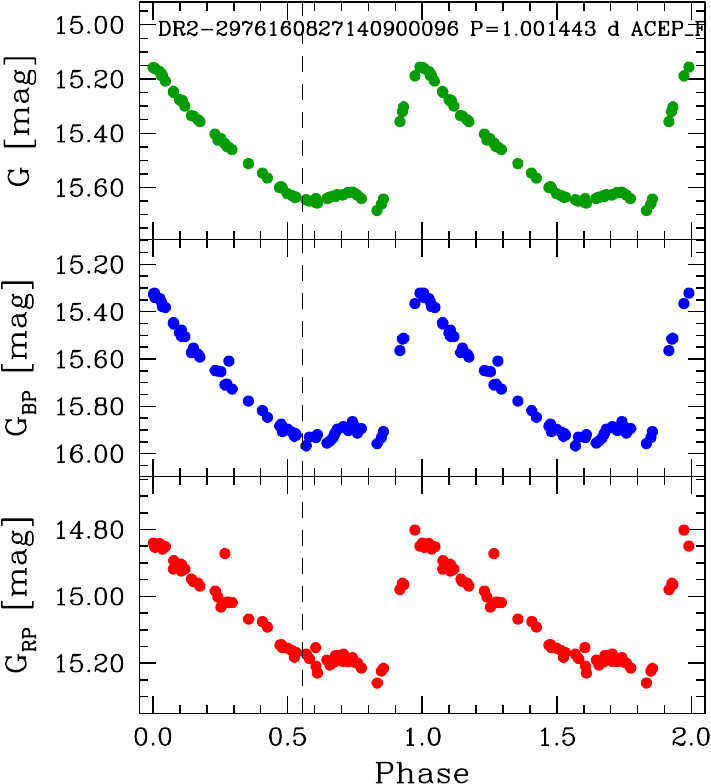}
}
\hbox{
\includegraphics[width=0.285\textwidth]{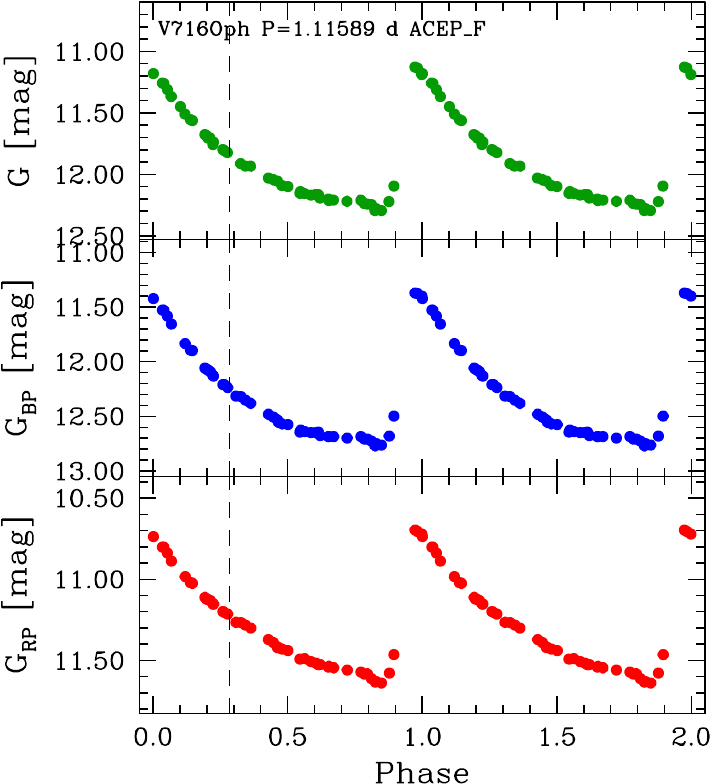}
\includegraphics[width=0.285\textwidth]{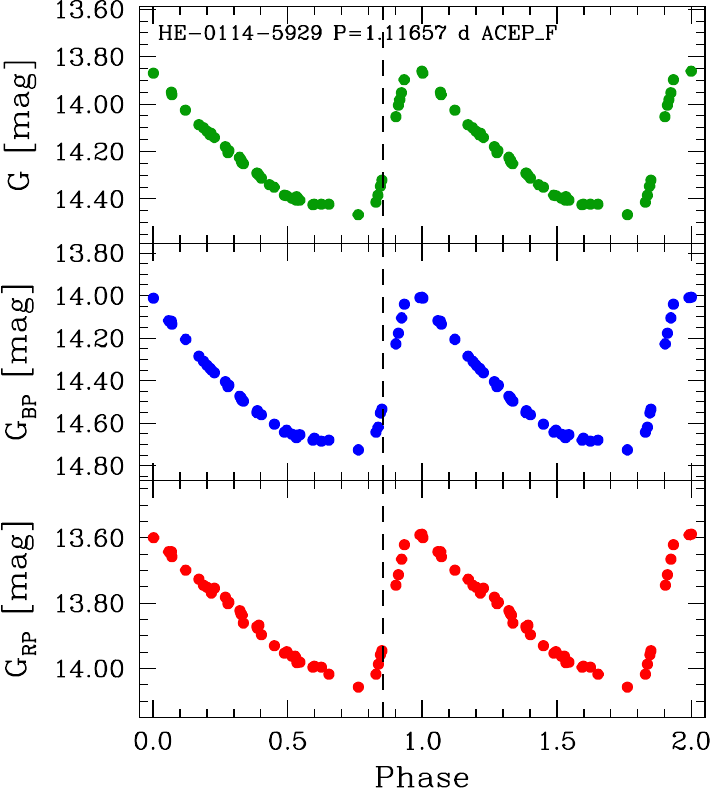}
\includegraphics[width=0.285\textwidth]{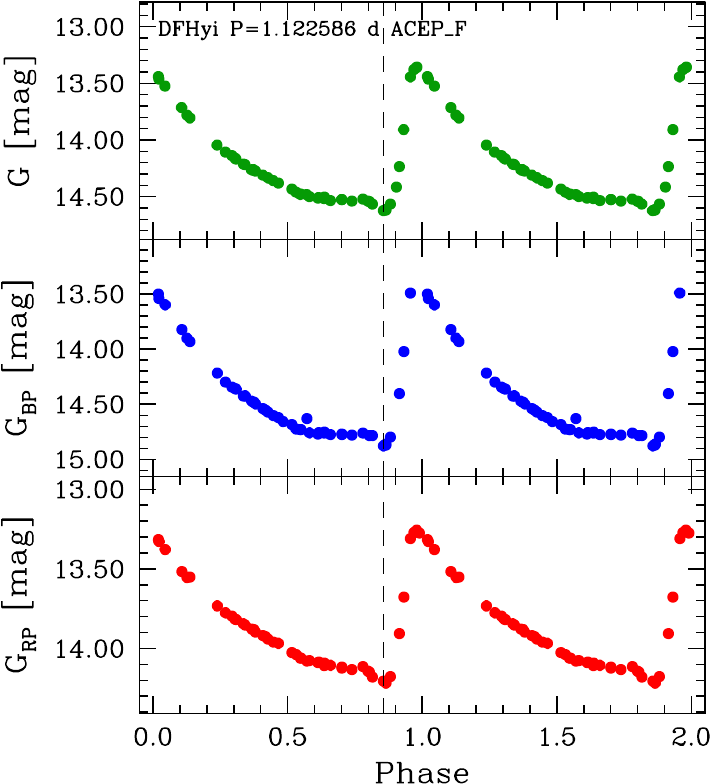}
}
\hbox{
\includegraphics[width=0.285\textwidth]{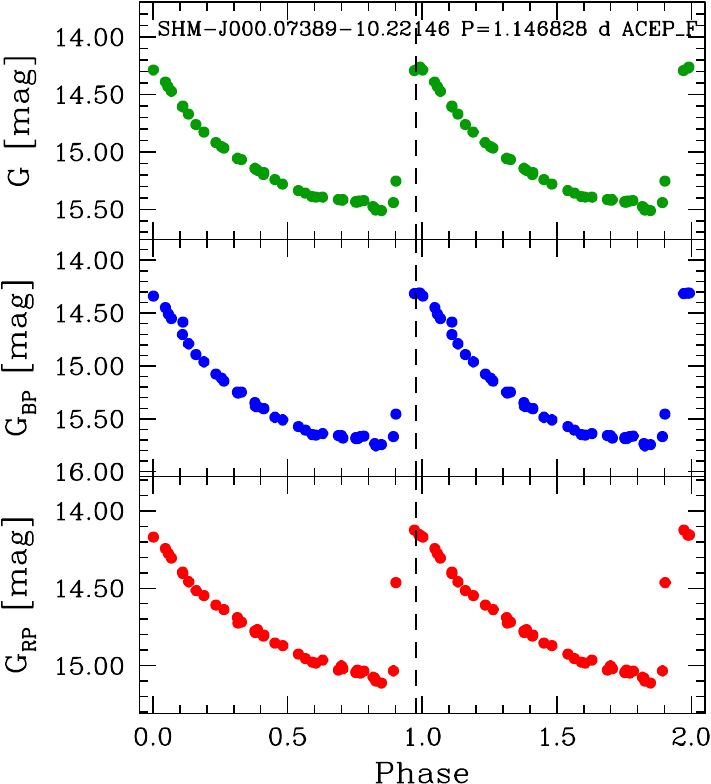}
\includegraphics[width=0.285\textwidth]{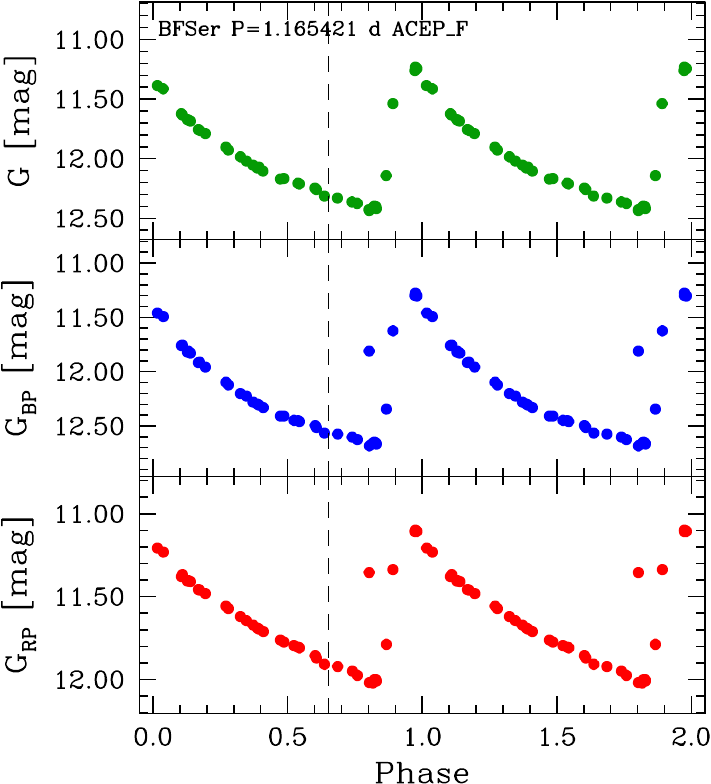}
\includegraphics[width=0.285\textwidth]{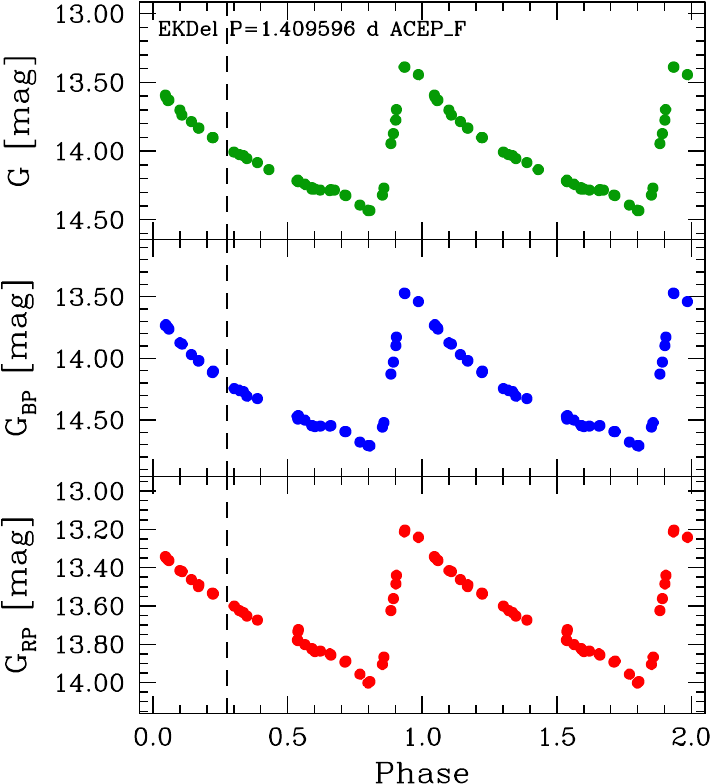}
}
\hbox{
\includegraphics[width=0.285\textwidth]{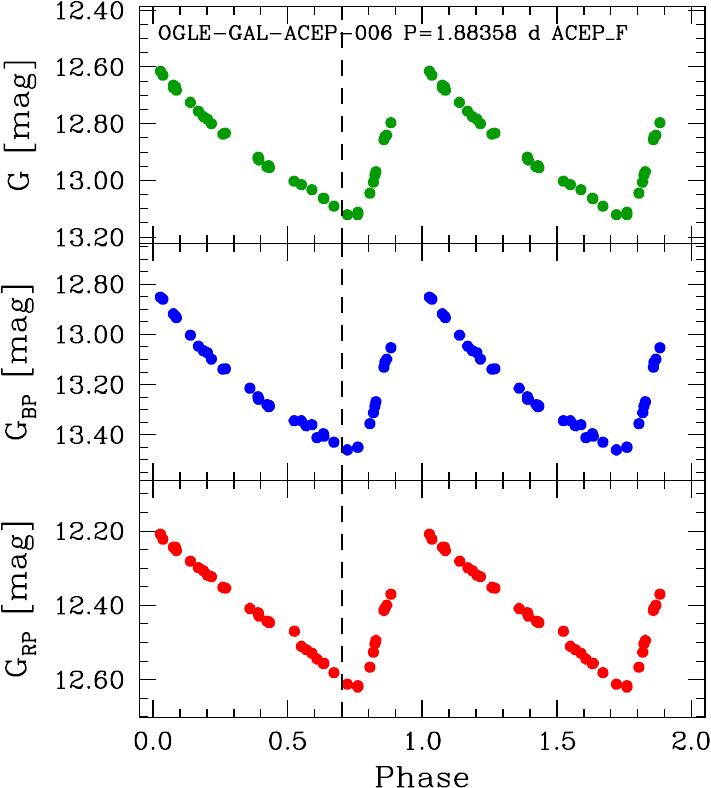}
\includegraphics[width=0.285\textwidth]{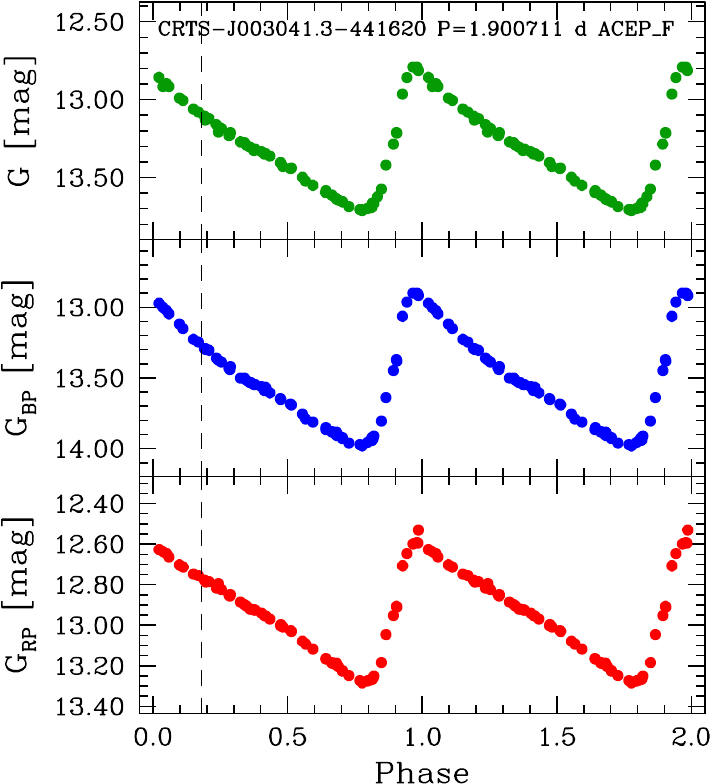}
\includegraphics[width=0.285\textwidth]{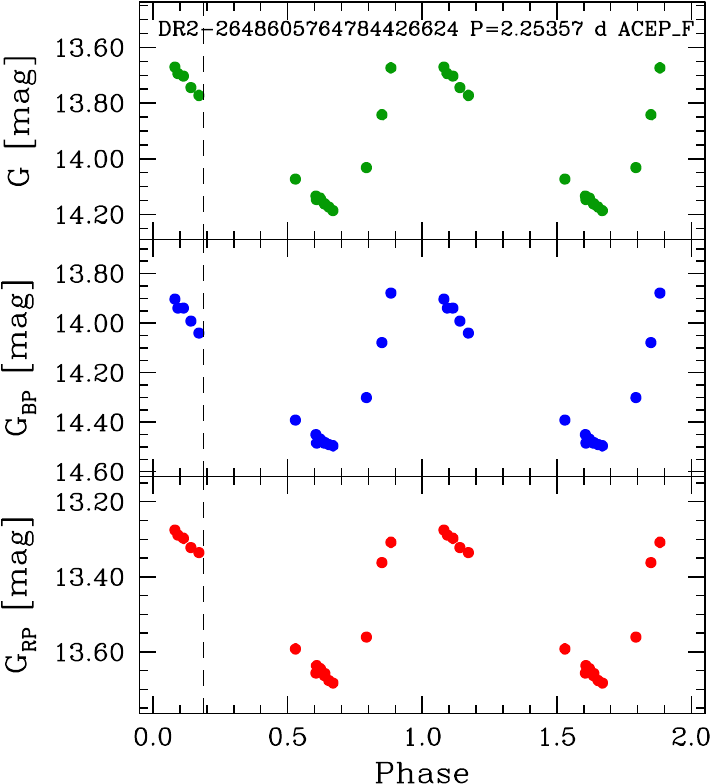}
}

}
\caption{Light curves for the target stars in the \gaia\ bands. The vertical line shows the phase corresponding to the spectroscopic observations. For BF Ser, EK Del and V716 Oph the phases have been taken from the literature. In particular, the line is only indicative for BF Ser and EK Del for which multiple epochs of observations are available (see text for details). 
              }
         \label{fig:lc}
   \end{figure*}

\FloatBarrier

\section{Abundances for the investigated ACEPs expressed as [X/Fe]}
\label{AppendixB}

\begin{table*}
\caption{Abundances for the stars analyzed in this work. All the data are expressed as [X/Fe] (dex).}
\label{tab:abundances_fe}
\footnotesize\setlength{\tabcolsep}{3pt}
\begin{tabular}{lccccccc}
\hline
\hline
  Star                       &         C         &       Na           &        Mg          &       Si          &        Ca        &          Sc       &          Ti         \\
\hline                                                                                                                                           
DR2 6498717390695909376      &         --      &    0.56 $\pm$ 0.19 &    0.54 $\pm$ 0.34 &         --      &  0.59 $\pm$ 0.19 &         --      &         --        \\ 
DR2 2976160827140900096      &         --      &    0.48 $\pm$ 0.18 &    0.29 $\pm$ 0.21 &         --      &  0.64 $\pm$ 0.22 &$-$0.16 $\pm$ 0.15 &   0.20 $\pm$ 0.22   \\
HE 2324-1255                 &         --      &    0.75 $\pm$ 0.20 &    0.30 $\pm$ 0.19 &$-$0.03 $\pm$ 0.19 &  0.41 $\pm$ 0.24 &$-$0.31 $\pm$ 0.23 &$-$0.11 $\pm$ 0.23   \\
V716 Oph$^{a}$               &   0.43 $\pm$ 0.28 &    0.36 $\pm$ 0.28 &    0.21 $\pm$ 0.28 &   0.14 $\pm$ 0.28 &  0.53 $\pm$ 0.28 &   0.23 $\pm$ 0.28 &   0.49 $\pm$ 0.28   \\
HE 0114-5929                 &         --      &    1.11 $\pm$ 0.15 &    0.27 $\pm$ 0.21 &         --      &  0.41 $\pm$ 0.21 &         --      &         --        \\
DF Hyi                       &         --      &    0.99 $\pm$ 0.16 &    0.46 $\pm$ 0.17 &$-$0.11 $\pm$ 0.19 &  0.49 $\pm$ 0.18 &$-$0.30 $\pm$ 0.18 &$-$0.04 $\pm$ 0.16   \\
SHM2017 J000.07389-10.22146  &         --      & $-$0.27 $\pm$ 0.11 & $-$0.42 $\pm$ 0.12 &   1.45 $\pm$ 0.11 &        --      &         --      &         --        \\
BF Ser$^{a}$                 &   0.13 $\pm$ 0.28 & $-$0.20 $\pm$ 0.28 &    0.31 $\pm$ 0.28 &         --      &  0.75 $\pm$ 0.28 &   0.20 $\pm$ 0.28 &   0.41 $\pm$ 0.28   \\
EK Del$^{b}$                 &$-$0.34 $\pm$ 0.28 &          --      &    0.01 $\pm$ 0.28 &   0.37 $\pm$ 0.28 &  0.30 $\pm$ 0.28 &         --      &   0.52 $\pm$ 0.28   \\
OGLE GAL-ACEP-006            &         --      &    0.86 $\pm$ 0.23 &    0.38 $\pm$ 0.19 &   0.50 $\pm$ 0.15 &  0.72 $\pm$ 0.16 &   0.01 $\pm$ 0.17 &   0.20 $\pm$ 0.19   \\
CRTS J003041.3-441620        &   0.42 $\pm$ 0.11 &    1.27 $\pm$ 0.14 &    0.38 $\pm$ 0.16 &$-$0.04 $\pm$ 0.11 &  0.48 $\pm$ 0.15 &$-$0.21 $\pm$ 0.17 &  -0.06 $\pm$ 0.18   \\
DR2 2648605764784426624      &         --      &    0.72 $\pm$ 0.22 &    0.13 $\pm$ 0.23 &   0.58 $\pm$ 0.21 &  0.50 $\pm$ 0.22 &   0.06 $\pm$ 0.19 &   0.31 $\pm$ 0.19   \\
\hline
 \multicolumn{8}{c}{Continued} \\
\hline  Star                   &          Cr         &       Ni       &         Y        &         Ba       \\
\hline                                                                                  
DR2 6498717390695909376        &  $-$0.10 $\pm$ 0.22 &        --      &         --      &         --      \\ 
DR2 2976160827140900096        &     0.10 $\pm$ 0.16 &        --      &         --      &$-$0.23 $\pm$ 0.19 \\ 
HE 2324-1255                   &  $-$0.09 $\pm$ 0.21 &        --      &         --      &$-$1.27 $\pm$ 0.21 \\ 
V716 Oph$^{a}$                 &     0.15 $\pm$ 0.28 &  0.32 $\pm$ 0.28 &   0.18 $\pm$ 0.28 &$-$0.05 $\pm$ 0.28 \\ 
HE 0114-5929                   &     0.01 $\pm$ 0.19 &        --      &         --      &         --      \\ 
DF Hyi                         &     0.22 $\pm$ 0.16 &  0.21 $\pm$ 0.14 &$-$0.13 $\pm$ 0.20 &$-$0.39 $\pm$ 0.20 \\ 
SHM2017 J000.07389-10.22146    &           --      &        --      &         --      &         --      \\ 
BF Ser$^{a}$                   &     0.26 $\pm$ 0.28 &        --      &   0.28 $\pm$ 0.28 &   0.20 $\pm$ 0.28 \\ 
EK Del$^{b}$                   &  $-$0.10 $\pm$ 0.28 &  0.26 $\pm$ 0.28 &   0.14 $\pm$ 0.28 &         --      \\ 
OGLE GAL-ACEP-006              &     0.05 $\pm$ 0.19 &        --      &$-$0.54 $\pm$ 0.16 &$-$0.08 $\pm$ 0.17 \\ 
CRTS J003041.3-441620          &     0.01 $\pm$ 0.18 &  0.20 $\pm$ 0.14 &$-$0.21 $\pm$ 0.18 &   0.44 $\pm$ 0.24 \\ 
DR2 2648605764784426624        &     0.15 $\pm$ 0.21 &  0.00 $\pm$ 0.23 &$-$0.43 $\pm$ 0.16 &$-$0.10 $\pm$ 0.20 \\ 
\hline 
\hline
\end{tabular}
\tablefoot{a=\citet{Kovtyukh2018}\\
b=\citet{Luck2011}} 
\end{table*}

\FloatBarrier

\section{Calculation of the integrals of motion}
\label{AppendixC}
To calculate the integral of motion we adopted the {\tt Galpy} package\footnote{http://github.com/jobovy/galpy} \citep{Bovy2015} adopting the {\tt McMillan17} potential \citep[][]{McMillan2017}. The calculation of the integral of motion requires positions, distances, proper motions and RVs. For the GGCs all these quantities are available on the {\it 4th version of our globular cluster database} \footnote{https://people.smp.uq.edu.au/HolgerBaumgardt/globular/} based on the papers by \citet{Vasiliev2021,Baumgardt2021}. Similarly, for the satellite of the MW we used the 
{\it Local Group and Nearby Dwarf Galaxies} database\footnote{https://www.cadc-ccda.hia-iha.nrc-cnrc.gc.ca/en/community/nearby/} based on the data published by \citet{McConnachie2020} to obtain the integral of motion for 54 Local Group members. \\
For the ACEPs, positions and proper motions are from \gaia\ DR3. In the same catalogue, we have found the average RVs (over many epochs) for 8 stars (including EK\,Del for which only the DR2 value is available), while for the remaining four (DR2\,2976160827140900096; DR2\, 6498717390695909376; SHM2017\,J000.07389-10.22146; HE\,2324-1255) we adopted the radial velocities measured from our spectra. As the RV of a pulsating star can vary up to a few tenths of km/s, these RVs are less precise than those present in the \gaia\ catalogues, but we do not expect that this occurrence impact significantly the derivation of the integral of motion. Calculating the distances of the ACEPs is a more complex task. First, we tried to use the \gaia\ EDR3-based distances from \citet{Bailer2021}, but the parallaxes were too small and the relative error on the parallaxes too large for many stars (see Table~\ref{tab:basicData}) so these distances substantially reflect the Galactic model used as prior. Therefore, we adopted a different approach. We assumed the $PW$ relation in the Gaia bands for the ACEP\_F in the LMC (the same data shown in Fig. ~\ref{fig:pw}) in the form published by \citet{Ripepi2023} but with the zero point calibrated in absolute terms using the geometric distance of the LMC based on eclipsing binaries \citep[18.477$\pm$0.026 mag, ][]{Pietrzynski2019}. The adopted $PW$ 
is $W_G=-1.687-3.080 \times \log P$, where $W_G=G-1.90 \times (G_{BP}-G_{RP})$ and $P$ is the period. Note that the period of the ACEP\_1O was fundamentalised using the relation $\log P_F=\log P_{1O}+0.13$ \citep[][]{Caputo2004}. The comparison of the absolute $W_G$ with the apparent one readily gives us the distances of the pulsators which are listed in Table~\ref{tab:basicData}. As a final note, we checked that these distances are in perfect agreement with those by \citet{Bailer2021} for the closest stars in our sample e.g. BF\,Ser and V716\,Oph.

\end{appendix}

\end{document}